\documentclass[a4paper,times]{emsart}

\usepackage{amsthm} 

\graphicspath{{figures/}}
\usepackage{algorithm,algorithmic}

\volumeyear{2026}
\volumenumber{XX}
\issuenumber{YY}
\journalname{Electromagnetic Science}
\startpage{1}
\DOI{123.123.123.123}
\journaltype{Original Article}

\title[Analytical Quantum Full-Wave Analysis of Few-Photon Transport Through a Waveguide Cavity Containing a Qubit]{Analytical Quantum Full-Wave Analysis of Few-Photon Transport Through a Superconducting Cavity Qubit}

\author{%
Soomin Moon\affilnums{1} and Thomas E. Roth\affilnums{1,2}
}

\affiliation{%
\affilnum{1}Elmore Family School of Electrical and Computer Engineering, Purdue University, West Lafayette, Indiana 47906, USA\\
\affilnum{2}Purdue Quantum Science and Engineering Institute, Purdue University, West Lafayette, Indiana 47907, USA 
}

\corrauth{Roth, Thomas E.}
\email{rothte@purdue.edu}

\receivetime{Received February 27, 2026}
\accepttime{Accepted XXXXXX YY, 2026}
\publishtime{Published XXXXXX YY, 2026}

\authorcopy{Soomin Moon and Thomas E. Roth}

\abstract{A promising way to scale up superconducting quantum computers is to link different devices together using propagating photons. Correspondingly, accurately modeling the quantum information transfer in such quantum interconnects is critical to advancing this emerging technology. To accomplish this, a full-wave quantum numerical model is essential for describing the few-photon transport characteristics of various components. Unfortunately, validating the accuracy of such numerical models remains a difficult challenge due to the lack of appropriate analytical solutions for standard component types. Recently, progress has been made on creating the first-ever analytical quantum full-wave solutions for a superconducting circuit quantum device. These efforts considered the case of two-photon transport through an empty rectangular waveguide cavity and the interactions of photons inside a closed rectangular waveguide cavity with a transmon qubit formed by a Josephson junction connected across the terminals of a small wire dipole antenna. Here, we advance these efforts by considering the one- and two-photon transport properties through a rectangular waveguide cavity containing a qubit in this form when the cavity is interfaced with via two coaxial ports. Such devices can be used in various ways for quantum interconnects, such as to form parts of a quantum memory or a photon source. We perform this analysis leveraging a quantum input-output theory formalism to derive the relevant single- and two-photon transport characteristics of interest. We then examine the signatures of the nonlinear quantum scattering effects in the good and bad cavity regimes of cavity quantum electrodynamics. In the future, these analytical results can be used to validate numerical full-wave quantum solvers for modeling quantum interconnects.}

\keywords{Circuit quantum electrodynamics, quantum electromagnetics, superconducting circuits, perturbation theory, electromagnetic theory.}

\begin{document}

\maketitle
\renewcommand{\thesection}{\Roman{section}}
\section{Introduction}
\label{sec:intro}
Quantum information processing is an emerging field that has the potential to revolutionize advancements in science and technology \cite{nielsen2016quantum}. One of the leading quantum processor platforms being explored is the circuit quantum electrodynamics (cQED) architecture \cite{krantz2019quantum,blais2021circuit}, which utilizes the interaction between microwave electromagnetic (EM) fields and superconducting circuits to generate and process quantum information. Although a leading platform with recent landmark advances on the road to quantum error correction \cite{google2025quantum}, significant design improvements need to be made to overcome the current engineering challenges faced in scaling these devices in size to have practical utility \cite{mohseni2024build}.  To address these requirements, engineers must explore new approaches in packaging, integration, and miniaturization methods, which consequently increase the complexity of the hardware \cite{brecht2016multilayer,huang2021microwave,kosen2022building,conner2021superconducting,rosenberg20173d,wang2022hexagonal,google2025quantum,karamlou2024probing}. To systematically engineer devices under these constraints, there is an increasing need for the development of high-fidelity numerical modeling tools that can analyze arbitrary cQED systems to continue advancing this field \cite{moon2024computational,mohseni2024build,Levenson-Falk_2025,shanto2024squadds,elkin2025opportunities}.

Unfortunately, there are few numerical methods available in this nascent field and they often suffer from efficiency issues \cite{moon2024computational}. For instance, one leading approach \cite{minev2021energy} took more than 28 hours to perform a simple parametric simulation of a notional two-qubit device \cite{Moon2024Analytical}. More specialized methods or alternative simulation strategies can help improve efficiency \cite{Moon2024Analytical,solgun2019simple,khan2024field,labarca2024toolbox,roth2024maxwell,elkin2025ims,khan2025incorporating}, but validating the accuracy and operating regimes where these methods are valid remains a significant difficulty. 
 
Generally, validating against an analytical solution is the first step in developing a new numerical method. In the area of cQED, this is challenging as most analytical solutions are only available for 1D systems \cite{blais2021circuit}. As an alternative route, comparing against other numerical methods is also challenging for cQED since there is a lack of efficient and fully-validated 3D numerical methods to compare against in many cases. Naturally, comparing against experimental measurements is also necessary. However, current devices still can contain too much uncertainty in their manufactured properties to achieve strong \textit{a priori} quantitative agreement with a numerical model for validation purposes. Further, the presence of noise and decoherence invariably exist in measurements due to various sources, but modeling these effects remains a significantly challenging task \cite{moon2024computational}. Correspondingly, a common intermediate step in developing a numerical method is to first consider ideal systems that omit noise and decoherence to verify a simpler model before continuing the development to incorporate these non-ideal effects. Given these considerations, it is clear that more analytical solutions that are 3D, full-wave, and capture the relevant quantum effects are needed for practical cQED devices.

In our prior work \cite{Moon2024Analytical}, we developed the first such analytical quantum full-wave solution for a geometry inspired by the experimental structure called a 3D transmon \cite{paik2011observation}. The 3D transmon is a system with a transmon qubit \cite{koch2007charge,roth2022transmon} formed by a planar dipole antenna supported on a substrate that is embedded inside the waveguide resonator. In contrast, our system considered a simplified geometry to make the analytical description of the system feasible. It consisted of a notional transmon qubit created by a wire dipole antenna inside a rectangular waveguide resonator, which is fed by multiple coaxial-shaped waveguide ports. Our analysis considered two kinds of analytical solutions to validate the modeling of quantum EM fields and qubit interactions. The first system considered a setup without the transmon qubit to analyze the photon transport properties through the cavity by using quantum input-output theory to formulate an analytical solution. These results were used to analyze the Hong-Ou-Mandel effect \cite{hong1987measurement,na2020classical} which is often used to qualitatively validate numerical methods for quantum EM methods \cite{na2020quantum,na2021diagonalization}. The second system considered a transmon qubit embedded in a closed cavity, which can be used to validate quantum modeling methods that consider field-qubit interactions.

While these two analytical solutions are useful for validating aspects of different numerical methods, it is desirable to have an analytical solution relevant to a complete device that consists of the cavity-qubit system that is interfaced with via coaxial ports. Such devices (especially planar variants) are relevant for building components of quantum interconnects, such as photon switches, quantum memories, or single-photon emitters and receivers \cite{lang2013correlations,kimble1998strong, bermel2006single, faraon2008coherent, monroe2014large, reinhard2012strongly, kannan2023demand, almanakly2025deterministic, brecht2016multilayer}, which are critical pieces needed to connect distant quantum processors together. Here, we create an analytical solution relevant to many such devices by building on the quantum input-output theory approaches of  \cite{shen2007strongly,shen2009theory, rephaeli2012few,fan2010input, xu2015input, xu2016fano} to analyze the single- and two-photon transport properties through our system of interest. 


In the earlier works of \cite{shen2007strongly, shen2009theory,fan2010input,rephaeli2012few, xu2015input, xu2016fano}, many artificial decompositions and transformations were defined to cast a problem into spaces of modes that only propagate in a single direction within a waveguide. Subsequent complex transformations were then defined to relate these one-way propagating solutions to the two-way propagating solutions relevant to an actual physical geometry. Other similar approaches have been taken in related works as well; e.g., \cite{oehri2015tunable, srinivasan2007mode}. Here, we show how we can directly consider our cavity-qubit system of interest without resorting to any such artificial decompositions and transformations, simplifying the derivation and making the physical manipulations clearer. Beyond this, we also leverage the classical EM theory techniques from \cite{Moon2024Analytical} to analytically evaluate all full-wave aspects of our system that are needed to compute the photon transport properties rather than treating all quantities in the expressions as fitting parameters, which is commonly done in these prior works. 


The remainder of the work is organized as follows. In Section \ref{sec:background}, we review the physical development of the Hamiltonian describing the complete cavity-qubit system coupled to port regions. Following this, we develop in Section \ref{sec:qiot} the quantum input-output theory equations of motion that will be solved in the one- and two-photon regimes analytically throughout this work. These solutions are developed for the single-photon case in Section \ref{sec:OnePhotonScattering} and for the two-photon case in Section \ref{sec:TwoPhotonScattering}. We then discuss in Section \ref{sec:g2} how the two-photon results can be used to compute the second-order correlation function that is typically measured in experiments to assess the properties of photon sources and other devices relevant to quantum interconnects. In Section \ref{sec:result}, we present numerical results relevant to the one- and two-photon transport properties of the cavity-qubit system when operated in the ``good'' and ``bad'' cavity regimes of cavity quantum electrodynamics (QED). We then draw final conclusions in Section \ref{sec:conclusion}. A number of appendices are also included to provide supporting details on the derivations of the main text.

\section{Background}
\label{sec:background}
\begin{figure}[t!]
    \centering
    \includegraphics[width=0.8\linewidth]      {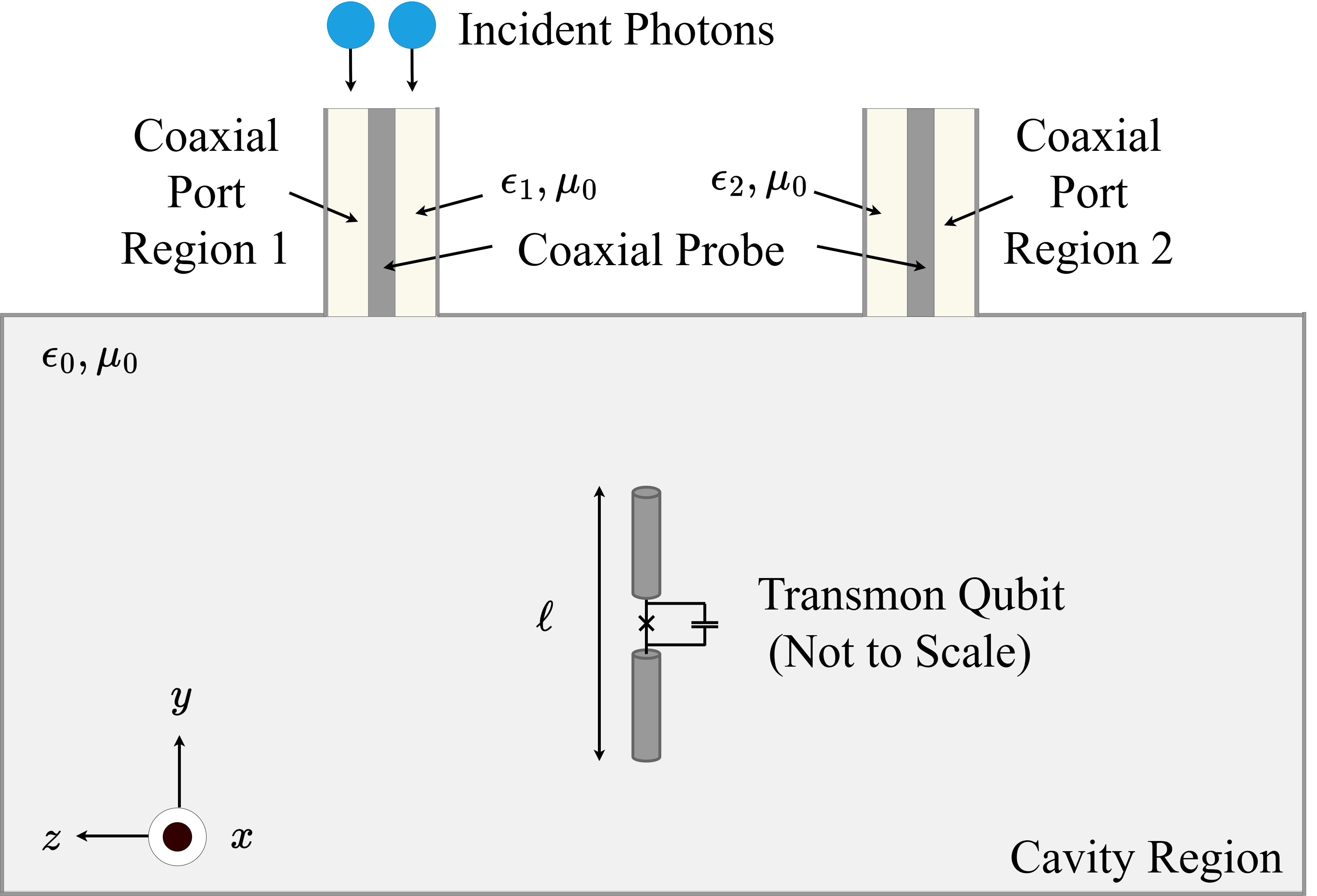}      
    \caption{Schematic of a transmon qubit inside a rectangular waveguide cavity that is directly connected to two coaxial ports that are infinitely long. In this work, we will analyze the scattering effects when one or two photons are incident on the cavity-qubit system.}
    \label{fig:cavity_qubit_port_geometries}
\end{figure}

Here, we will review the formulation of the Hamiltonian appropriate for describing the system shown in Fig. \ref{fig:cavity_qubit_port_geometries} following the macroscopic cQED formalism of \cite{roth2021macroscopic,Moon2024Analytical}. In this approach, the EM fields are quantized in a full-wave manner within the context of macroscopic quantum electrodynamics \cite{scheel2008macroscopic} where dielectric materials are considered in terms of macroscopic quantities like the permittivity rather than through microscopic descriptions (introductions to this kind of field quantization can be found in \cite{chew2016quantum,chew2016quantum2,chew2021qme-made-simple}). The superconducting qubits are also treated in a macroscopic fashion, with the collective effects of their underlying superconducting circuit leading to an effective lumped-element description. While additional superconducting effects can be critical in accurately modeling typical experiments \cite{elkin2025opportunities}, these are less prominent for the system of Fig. \ref{fig:cavity_qubit_port_geometries} due to the large thicknesses of the materials and large mode volumes \cite{nigg2012black}, and so will be neglected for simplicity by approximating the superconducting materials as perfect electric conductors (PECs).

Now, the total Hamiltonian describing the system shown in Fig. \ref{fig:cavity_qubit_port_geometries} can be decomposed into three separate sub-systems and the corresponding interactions between them, with the detailed developments of each of these pieces available in \cite{Moon2024Analytical}. Overall, the total Hamiltonian is given by
\begin{align}
    \hat{H} = \hat{H}_\mathrm{C} + \hat{H}_\mathrm{T} + \hat{H}_\mathrm{CT} + \hat{H}_\mathrm{P} +\hat{H}_\mathrm{CP},
    \label{eq:total_Hamiltonian}
\end{align}
where the first three terms characterize the Hamiltonians of the cavity EM fields, the transmon qubit, and the interactions between them. The final terms characterize the Hamiltonian of the EM fields in the coaxial port regions and the interactions of these with the cavity fields.

Beginning with the cavity region, we have that
\begin{align}
    \hat{H}_\mathrm{C} = \frac{1}{2} \iiint \big( \epsilon\hat{\mathbf{E}}^{2}_c + \mu_0\hat{\mathbf{H}}^{2}_c \big) \mathrm{d}V,
    \label{eq:Hamiltonian_Cavity_field}
\end{align}
where the electric and magnetic field operators are given by
\begin{align}
    \hat{\mathbf{E}}_{c}(\mathbf{r},t)  = \sqrt{\frac{\hbar\omega_{c}}{2\epsilon_{0}}} \big(\hat{a}_c(t) + \hat{a}^{\dag}_{c}(t) \big) \mathbf{E}_{c}(\mathbf{r}),
    \label{eq:eigenmode expansion - Efield - cavity1}
\end{align}
\begin{align}
    \hat{\mathbf{H}}_{c}(\mathbf{r},t)  = -i \sqrt{\frac{\hbar\omega_{c}}{2\mu_{0}}}\big(\hat{a}_c(t) - \hat{a}^{\dag}_c(t) \big)\mathbf{H}_{c}(\mathbf{r}).
    \label{eq:eigenmode expansion - Hfield - cavity1}
\end{align}
As in \cite{Moon2024Analytical}, we only consider a single cavity field mode to keep the analytical treatments more manageable. For the purposes of this work, this will be the fundamental cavity field mode; however, higher-order modes could also be considered so long as the one considered has a well-separated resonant frequency from nearby field modes (a condition required by the quantum input-output theory approach used in Section \ref{sec:qiot}). In these expressions, the cavity resonant frequency is given by $\omega_c$ and the spatial mode distributions of the electric and magnetic fields are $\mathbf{E}_c(\mathbf{r})$ and $\mathbf{H}_c(\mathbf{r})$, respectively. Finally, $\hat{a}_c$ and $\hat{a}^\dagger_c$ are the bosonic creation and annihilation operators of the cavity field mode, which satisfy the commutation relation $[\hat{a}_c(t), \hat{a}_c^{\dagger}(t)] = 1$.

Next, the free transmon Hamiltonian is given by
\begin{align}
\hat{H}_\mathrm{T} = 4E_C\hat{n}^2 - E_J\cos\hat{\varphi},
\label{eq:transmon_Hamiltonian}
\end{align}
where $E_C$ and $E_J$ are the charging and Josephson energies of the qubit \cite{koch2007charge,roth2022transmon}. Further, $\hat{n}$ and $\hat{\varphi}$ are the qubit charge and phase operators. For our purposes of developing an analytical solution of the few-photon transport through the cavity containing a qubit of this kind, we will only consider the presence of the first two energy levels (denoted by $|\mathrm{g}\rangle$ and $|\mathrm{e}\rangle$) of the transmon in our model. In this case, we can re-express the transmon Hamiltonian within this basis as
\begin{align}
    \hat{H}_\mathrm{T} = \frac{1}{2} \hbar\omega_q \hat{\sigma}_z,
    \label{eq:transmon_Hamiltonian2}
\end{align}
where $\hat{\sigma}_z = |\mathrm{e}\rangle\langle \mathrm{e}| - |\mathrm{g}\rangle\langle \mathrm{g}|$ and $\omega_q$ is the transition frequency between the energy levels $|\mathrm{g}\rangle$ and $|\mathrm{e}\rangle$.

The interaction Hamiltonian between the qubit and cavity field is
\begin{align}
\hat{H}_\mathrm{CT} = 2e \hat{n} \int\hat{\mathbf{E}}_{c}\cdot\mathbf{d}(\mathbf{r}) \mathrm{d}\mathbf{r},
\label{eq:Hamiltonian_int_CT}
\end{align}
where $e$ is the elementary charge and $\mathbf{d}$ parameterizes a line integration path so that its integral with $\hat{\mathbf{E}}_c$ computes the voltage seen by the Josephson junction in the transmon qubit \cite{roth2021macroscopic,Moon2024Analytical}. Using the expression for $\hat{\mathbf{E}}_c$ from (\ref{eq:eigenmode expansion - Efield - cavity1}) and further expressing the qubit operators in terms of the basis of $|\mathrm{g}\rangle$ and $|\mathrm{e}\rangle$, (\ref{eq:Hamiltonian_int_CT}) becomes
\begin{align}
    \hat{H}_\mathrm{CT} = \hbar g \big( \hat{a}_c^{\dagger}\hat{\sigma}_{-} + \hat{a}_c\hat{\sigma}_{+}\big),
    \label{eq:H_CT_2}
\end{align}
where $\hat{\sigma}_+ = |\mathrm{e}\rangle\langle \mathrm{g}|$, $\hat{\sigma}_- = |\mathrm{g}\rangle\langle \mathrm{e}|$, and 
\begin{align}
g = 2e \langle \mathrm{g} | \hat{n} | \mathrm{e} \rangle  \sqrt{\frac{\omega_c}{2\epsilon_{0}\hbar} }\int{\mathbf{E}_c(\mathbf{r}})\cdot \mathbf{d}(\mathbf{r}) \mathrm{d}\mathbf{r}.
\label{eq:coupling_strength_g}
\end{align}
Note that in arriving at (\ref{eq:H_CT_2}) we have made a rotating wave approximation (RWA) to drop fast oscillating terms that would have a negligible impact on the system dynamics for the configurations considered here \cite{walls2007quantum}. We have further assumed that the phases of the qubit states have been set appropriately so that $\langle \mathrm{g} | \hat{n} | \mathrm{e} \rangle$ is real.

The remaining terms deal with the port sub-systems. The free Hamiltonian of the ports is
\begin{align}
\hat{H}_\mathrm{P} = \sum_p \frac{1}{2} \iiint \big( \epsilon_p\hat{\mathbf{E}}^{2}_p + \mu_0\hat{\mathbf{H}}^{2}_p \big)\mathrm{d}V,
\label{eq:Hamiltonian_Port_field}
\end{align}
where $\epsilon_p$ is the permittivity in the $p$th coaxial subdomain. For ports of semi-infinite length, the EM fields are represented as a continuum of modes as
\begin{multline}
    \hat{\mathbf{E}}_{p}(\mathbf{r},t)  =   \int_{0}^{\infty} \sqrt{\frac{\hbar u}{2\epsilon_{0}}}  \big(\hat{a}_{ p}(t,u)\! +\! \hat{a}^{\dag}_{p}(t,u)\big) \\ \times\mathbf{E}_{ p}(u,\mathbf{r}) \, \mathrm{d} u,
    \label{eq:eigenmode expansion - Efield - port}
\end{multline}
\begin{multline}
    \hat{\mathbf{H}}_{p}(\mathbf{r},t)  =  -i \int_{0}^{\infty} \sqrt{\frac{\hbar u}{2\mu_{0}}}  \big(\hat{a}_{ p}(t,u)\! -\! \hat{a}^{\dag}_{ p}(t,u)\big) \\ \times\mathbf{H}_{ p}(u,\mathbf{r}) \, \mathrm{d}u,
    \label{eq:eigenmode expansion - Hfield - port}
\end{multline}
where $u$ is the frequency of the mode and only the transverse electromagnetic mode of the coaxial transmission lines are considered. In the continuum case, the bosonic creation and annihilation operators have commutation relation
\begin{align}
    [\hat{a}_{ p_1}(t,u) , \hat{a}^\dagger_{p_2}(t,v)] =  \delta_{p_1 p_2} \delta(u - v).
    \label{eq:port_field_commutation}
\end{align}

Finally, the interaction Hamiltonian describing the coupling between the cavity and port fields comes from the mode matching quantization approach detailed in \cite{roth2021macroscopic,Moon2024Analytical} and is given by
\begin{align}
    \hat{H}_\mathrm{CP} = - \sum_p \iint \hat{\mathbf{F}}_c \cdot \big(  \hat{\mathbf{E}}_{ p} \times \tilde{n}_p  \big)  \mathrm{d}S.
    \label{eq:Hamiltonian_cavity_port_interaction}
\end{align}
Here, the surface integral takes place in the annular region at the interface between the coaxial ports and the cavity and $\tilde{n}_p$ is the unit normal vector pointing into the cavity. Further, the electric vector potential in the cavity region is
\begin{align}
    \hat{\mathbf{F}}_c(\mathbf{r},t) = -\sqrt{\frac{\hbar}{2\omega_c \mu_0}} \big( \hat{a}_c(t) + \hat{a}^\dagger_c(t) \big) \mathbf{H}_c(\mathbf{r}).
\end{align}
Overall, this interaction can be viewed as coming from an effective magnetic current density at the coaxial port aperture interacting with the cavity fields.

Putting all these results together and leveraging the orthonormality properties of the various EM field modes, the total Hamiltonian of (\ref{eq:total_Hamiltonian}) becomes
\begin{multline}
    \hat{H}_\mathrm{tot} \!=\!  \hbar\omega_c \hat{a}^\dagger_c \hat{a}_c + 
 \frac{1}{2}\hbar \omega_{q} \hat{\sigma}_{z}  +
\hbar g\big( \hat{a}_c^{\dagger}\hat{\sigma}_{-} + \hat{a}_c\hat{\sigma}_{+}\big) \\+  \sum_{p} \int_0^\infty \hbar u \hat{a}^\dagger_{ p} (u) \hat{a}_{ p}(u) \mathrm{d}u \, \\+
     \sum_{ p } \int_{0}^{\infty} \!\! \hbar g_{ p}  \big(\hat{a}_{c}^{\dag}\hat{a}_{ p}(u) \!+\! \hat{a}_{c} \hat{a}_{ p}^{\dag}(u) \big)  \mathrm{d} u,
     \label{eq:Hamiltonian_empty_cavity_simplified}
\end{multline}
where
\begin{align}
    g_{p} = \iint \frac{c_0}{2}\sqrt{\frac{\omega_{p}}{\omega_c}} \big[\mathbf{H}_{c} \cdot (\mathbf{E}_{p} \times \tilde{n}_p) \big] \mathrm{d}S
    \label{eq:coupling-strength}
\end{align}
and we have suppressed the dependence on time for the various operators to simplify the notation. We have also made another RWA to simplify the interaction between the cavity and port field operators.

\section{Quantum Input-Output Theory}
\label{sec:qiot}
We now turn to establishing the equations of motion associated with the Hamiltonian of (\ref{eq:Hamiltonian_empty_cavity_simplified}) within the framework of quantum input-output theory \cite{walls2007quantum}. To begin, we first evaluate the Heisenberg equations of motion for the annihilation operators of the qubit, cavity fields, and port fields. These are
\begin{align}
    \partial_t \hat{\sigma}_{-}(t) = -i\omega_q \hat{\sigma}_{-}(t) + ig\hat{\sigma}_{z}(t)\hat{a}_c(t),
\label{eq:Fan_EoM_qubit_lowering_operator}
\end{align}
\begin{multline}
    \partial_t \hat{a}_c(t) = -i\omega_c \hat{a}_c(t) - ig\hat{\sigma}_{-}(t) \\ -\sum_{p} i \int  g_p\hat{a}_{p}(t,u) \mathrm{d} u ,
\label{eq:Fan_EoM_c}    
\end{multline}
\begin{align}
    \partial_t \hat{a}_{p}(t,u) = -iu\hat{a}_{p}(t,u) -ig_p\hat{a}_c(t).
\label{eq:Fan_EoM_ak}
\end{align}

The quantum input-output theory approach to analyzing photon transport properties involves integrating the port equations of motion in terms of initial and final conditions that occur well before and well after the interactions with the cavity system have occurred to formally solve for $\hat{a}_{p}(t,u)$ \cite{walls2007quantum,Moon2024Analytical}. As part of this process, we can define the input and output field operators in the ports as 
\begin{align}
    \hat{a}_{\mathrm{in},p}(t) = \frac{1}{\sqrt{2 \pi}}\int_{-\infty}^{\infty}  e^{-iu(t-t_0)}\hat{a}_{p}(t_0,u)  \mathrm{d} u,
	\label{eq:ain}
\end{align}
\begin{align}
    \hat{a}_{\mathrm{out},p}(t)  =   \frac{1}{\sqrt{2 \pi}}\int_{-\infty}^{\infty}   e^{-iu(t-t_1)}\hat{a}_{p}(t_1,u)  \mathrm{d} u,
	\label{eq:aout}
\end{align}
where the reference time $t_0$ ($t_1$) is well before (after) the interaction occurs. These input and output operators satisfy the commutation relationship
\begin{multline}
    [\hat{a}_{\mathrm{in},p}(t), \hat{a}^{\dagger}_{\mathrm{in},b}(t^{\prime})] = [\hat{a}_{\mathrm{out},p}(t), \hat{a}^{\dagger}_{\mathrm{out},b}(t^{\prime})] \\ = \delta_{p,b}\delta(t-t^{\prime}),
\end{multline}
which is proved in Appendix \ref{sec:QIOT_operator_commutation} and will be useful in our later solution process. 

The quantum input-output theory approach continues by substituting the formal solution of $\hat{a}_{p}(t,u)$ into the cavity equation of motion given by (\ref{eq:Fan_EoM_c}).
By assuming a Markovian approximation that $g_p$ does not vary significantly over the bandwidth of interest (nominally, the linewidth of the cavity), the quantum input-output equation of motion for the cavity in terms of the initial time condition can be found to be (with the full details available in \cite{Moon2024Analytical})
\begin{multline}
    \partial_t \hat{a}_c(t) = \bigg(-i\omega_c - \sum_{p}\frac{\kappa_p}{2} \bigg) \hat{a}_c(t) \\  - ig\hat{\sigma}_{-}(t) -\sum_{p} i\sqrt{\kappa_p} \hat{a}_{\mathrm{in},p}(t) ,
\label{eq:Fan_EoM_c_input_output_QIOT}
\end{multline}
where $\kappa_p \equiv 2\pi g_{p}^2 $ is the cavity decay rate through port $p$ \cite{walls2007quantum}. A similar equation can also be derived in terms of the final time condition. Subtracting these cavity equations of motion in terms of initial and final time conditions from each other, we can arrive at a local port quantum input-output theory ``boundary condition'' given by \cite{walls2007quantum}
\begin{align}
    \hat{a}_{\mathrm{out},p}(t) = \hat{a}_{\mathrm{in},p}(t) -i\sqrt{\kappa_p}\hat{a}_c(t).
\label{eq:Fan_input_output_relationship}
\end{align}
This equation, along with (\ref{eq:Fan_EoM_qubit_lowering_operator}) and (\ref{eq:Fan_EoM_c_input_output_QIOT}), will allow us to analyze the photon transport properties through the cavity-qubit system of Fig. \ref{fig:cavity_qubit_port_geometries}.

\section{Single-Photon Transport}
\label{sec:OnePhotonScattering}
With the basic quantum input-output theory equations developed for our case of interest, we can now analyze the single-photon transport properties through the cavity-qubit system. Such transport properties are typically expressed in the frequency domain through the relationship of freely-propagating input and output photon states in the port regions. These states are defined through the Fourier transform of $\hat{a}_{\mathrm{in},p}(t)$ and $\hat{a}_{\mathrm{out},p}(t)$. Correspondingly, we note that these frequency domain operators are given by
\begin{align}
    \hat{a}_{\mathrm{in},p}(u) = \frac{1}{\sqrt{2 \pi}}\int_{-\infty}^{\infty}  \hat{a}_{\mathrm{in},p}(t) e^{i u t} \mathrm{d} t ,
	\label{eq:ain_freq_FT}
\end{align}
with the same transform to define $\hat{a}_{\mathrm{out},p}(u)$ as well. Now, we can define an input field mode of a single photon of frequency $u$ that is incident through Port $p$ as
\begin{align}
    |u^{+}\rangle_{p} = \hat{a}^{\dagger}_{\mathrm{in},p}(u)| 0 \rangle,
    \label{eq:input_scattering_state}
\end{align}
where the superscript ``$+$'' denotes the input state propagating toward the cavity-qubit system. Note that throughout this work, we use $| 0 \rangle$ to represent the global ground state of the entire system of ports, cavity, and qubit. Similarly, the output field mode of a single photon of frequency $v$ that exits through Port $b$ corresponds to 
\begin{align}
    |v^{-}\rangle_{b} = \hat{a}^{\dagger}_{\mathrm{out},b}(v)| 0 \rangle,
    \label{eq:output_scattering_state}
\end{align}
where the superscript ``$-$'' denotes the output state propagating away from the cavity-qubit system. 

The single-photon transport properties are then given in terms of the inner product between the various kinds of input and output states relevant to a particular system. The general expression in the single-photon case is then
\begin{align}
{}_{b}\langle v^{-}|u^{+} \rangle_{p} = \langle 0|\hat{a}_{\mathrm{out},b_{}}(v)\hat{a}^{\dagger}_{\mathrm{in},p_{}}(u)| 0 \rangle,
\label{eq:ScatteringMatrix_Single}
\end{align}
which is also sometimes referred to as an element of the scattering matrix for the system \cite{fan2010input,rephaeli2012few,xu2015input}. A convenient way to derive the result of such inner products is to use the frequency domain version of (\ref{eq:Fan_input_output_relationship}) to re-express $\hat{a}_{\mathrm{out},b_{}}(v)$ in terms of the input field operator and cavity operators. Substituting this frequency domain version of (\ref{eq:Fan_input_output_relationship}) into (\ref{eq:ScatteringMatrix_Single}), we get
\begin{multline}
    {}_{b_{}} \langle v^{-}| u^{+} \rangle_{p_{}} =  \langle 0|\hat{a}_{\mathrm{in},b_{}}(v)\hat{a}^{\dagger}_{\mathrm{in},p_{}}(u)| 0 \rangle \\ -i\sqrt{\kappa_{b_{}}}\langle 0|\hat{a}_c(v) \hat{a}^{\dagger}_{\mathrm{in},p_{}}(u)|0 \rangle_{p_{}}.
\label{eq:ScatteringMatrix_Single_QIOTBC_1}
\end{multline}
To simplify this, we can use the frequency domain version of the commutation relations of the input operators given by
\begin{align}
    [\hat{a}_{\mathrm{in},p}(u), \hat{a}^{\dagger}_{\mathrm{in},b}(v)] 
 = \delta(u - v) \delta_{b,p},
\end{align}
which is proved in Appendix \ref{sec:QIOT_operator_commutation}. We then see that (\ref{eq:ScatteringMatrix_Single_QIOTBC_1}) becomes
\begin{align}
    {}_{b_{}} \langle v^{-}| u^{+} \rangle_{p_{}} = \delta(u-v)\delta_{b_{}, p_{}} -i\sqrt{\kappa_{b_{}}} \langle 0|\hat{a}_c(v)| u^{+} \rangle_{p_{}}.
\label{eq:ScatteringMatrix_Single_QIOTBC}
\end{align}
Hence, to determine the single-photon transport properties we must solve for $\langle 0|\hat{a}_c(v)| u^{+} \rangle_{p_{}}$.

Considering this, the overall solution strategy followed is to analyze the relevant matrix elements of the equations of motion presented in Section \ref{sec:qiot} to find a closed system to solve for $\langle 0|\hat{a}_c(v)| u^{+} \rangle_{p_{}}$. Focusing first on the relevant cavity and qubit equations of motion, we have that (\ref{eq:Fan_EoM_qubit_lowering_operator}) and (\ref{eq:Fan_EoM_c_input_output_QIOT}) become
\begin{multline}
    \partial_t \langle 0 |\hat{\sigma}_{-}(t)|u^{+} \rangle_{p}  = -i\omega_q \langle 0 |\hat{\sigma}_{-}(t)|u^{+} \rangle_{p} \\ + ig\langle 0 |\hat{\sigma}_{z}(t)\hat{a}_c(t)|u^{+} \rangle_{p},
\label{eq:Fan_EoM_qubit_lowering_operator_matrix}
\end{multline}
\begin{multline}
    \partial_t \langle 0 |\hat{a}_c(t)|u^{+} \rangle_{p} = \bigg(-i\omega_c - \sum_p\frac{\kappa_p}{2} \bigg) \langle 0 |\hat{a}_c(t)|u^{+} \rangle_{p} \\ -\sum_p i\sqrt{\kappa_p} \langle 0 |\hat{a}_{\mathrm{in},p}(t) |u^{+} \rangle_{p} - ig \langle 0|\hat{\sigma}_{-}(t)|u^{+}\rangle_{p}.
\label{eq:Fan_EoM_c_input_output_QIOT_matrix}
\end{multline}
The final term in (\ref{eq:Fan_EoM_qubit_lowering_operator_matrix}) poses difficulties in formulating a closed system. However, for typical single-photon scattering conditions, this can be simplified by assuming that the qubit remains predominantly in its ground state throughout the entire scattering process. This is typically a valid assumption when the pulse width of the photon is much greater than the spontaneous emission lifetime of the qubit \cite{rephaeli2012few, rephaeli2010full, fan2010input,waks2006dipole, roy2017colloquium}, which is the relevant case here where we focus on the steady-state input and output fields at (approximately) single frequency excitations. Considering this simplification, we can note that $\langle 0 |\hat{\sigma}_{z}(t)\hat{a}_c(t)|u^{+} \rangle_{p} \approx -\langle 0 |\hat{a}_c(t)|u^{+} \rangle_{p}$ because the states involved only have $\sigma_z(t)$ acting on the qubit ground state $|\mathrm{g}\rangle$ throughout the scattering process. Then, (\ref{eq:Fan_EoM_qubit_lowering_operator_matrix}) becomes
\begin{multline}
    \partial_t \langle 0 |\hat{\sigma}_{-}(t)|u^{+} \rangle_{p}  = -i\omega_q \langle 0 |\hat{\sigma}_{-}(t)|u^{+} \rangle_{p} \\ - ig\langle 0 |\hat{a}_c(t)|u^{+} \rangle_{p}.
\label{eq:Fan_EoM_qubit_lowering_operator_matrix_assump}
\end{multline}

To solve for $\langle 0|\hat{a}_c(v)| u^{+} \rangle_{p_{}}$, we take the Fourier transform of (\ref{eq:Fan_EoM_c_input_output_QIOT_matrix}) and (\ref{eq:Fan_EoM_qubit_lowering_operator_matrix_assump}), which results in
\begin{multline}
    -i v\langle 0 |\hat{a}_c(v)|u^{+} \rangle_{p} = -i\bigg(\omega_c -i \sum_p\frac{\kappa_p}{2} \bigg) \langle 0 |\hat{a}_c(v)|u^{+} \rangle_{p} \\ -\sum_p i\sqrt{\kappa_p} \langle 0 |\hat{a}_{\mathrm{in},p}(v) |u^{+} \rangle_{p}  - ig \langle 0|\hat{\sigma}_{-}(v)|u^{+}\rangle_{p},
\label{eq:Fan_EoM_c_input_output_QIOT_matrix_Fourier}
\end{multline}
\begin{multline}
    -i v\langle 0 |\hat{\sigma}_{-}(v)|u^{+} \rangle_{p}  = -i\omega_q \langle 0 |\hat{\sigma}_{-}(v)|u^{+} \rangle_{p} \\ - ig\langle 0 |\hat{a}_c(v)|u^{+} \rangle_{p}.
\label{eq:Fan_EoM_qubit_lowering_operator_matrix_assump_Fourier}
\end{multline}
Noting that
\begin{align}
    \langle 0 |\hat{a}_{\mathrm{in},p}(v) |u^{+} \rangle_{p} = \delta(u-v),
\end{align}
we can solve (\ref{eq:Fan_EoM_c_input_output_QIOT_matrix_Fourier}) and (\ref{eq:Fan_EoM_qubit_lowering_operator_matrix_assump_Fourier}) together to get
\begin{multline}
    \langle 0 |\hat{\sigma}_{-}(v)|u^{+} \rangle_{p}  
    \\ = \frac{g\sqrt{\kappa_{p}}}{ \big[ (v-\omega_c) +i\sum_p i\frac{\kappa_p}{2} \big](v-\omega_q) - g^2 } \delta{(u-v)} 
    \\ = S_{q,p}(v) \delta{(u-v)},
\label{eq:Fan_EoM_qubit_lowering_operator_matrix_assump_Fourier_result}
\end{multline}
\begin{multline}
    \langle 0 |\hat{a}_c(v)|u^{+} \rangle_{p} \\ 
    = \frac{\sqrt{ \kappa_{p}}(v-\omega_q) }{ \big[ (v-\omega_c) +i\sum_p i\frac{\kappa_p}{2} \big](v-\omega_q) - g^2 } \delta{(u-v)}  
    \\ = S_{c,p}(v) \delta{(u-v)}.     
\label{eq:Fan_EoM_c_input_output_QIOT_matrix_Fourier_result}
\end{multline}

Finally, substituting $\langle 0 |\hat{a}_c(v)|u^{+} \rangle_{p}$ into (\ref{eq:ScatteringMatrix_Single_QIOTBC}) provides the desired single-photon transport properties. If we specifically consider the case where the single photon is incident from Port 1, we have that the reflection properties are
\begin{multline}
    \! \! \! {}_{1}\langle v^{-}| u^{+} \rangle_{1} 
     = \frac{{ \big[(v-\omega_c) + \big(-i\frac{\kappa_{1}}{2} + i\frac{\kappa_{2}}{2}  \big)\big](v-\omega_q) - g^2 } }{ \big[(v-\omega_c) +\sum_p i\frac{\kappa_p}{2} \big)\big](v-\omega_q) - g^2 } \\ \times \delta{(u-v)}  = r_{1}(v) \delta{(u-v)}
\label{eq:Smatrix_2port_SinglePhoton_reflection}
\end{multline}
and the transmission properties to Port 2 are
\begin{multline}
    {}_{2}\langle v^{-}| u^{+} \rangle_{1} 
    =  \frac{-i\sqrt{\kappa_{1} \kappa_{2}}(v-\omega_q) }{ \big[ (v-\omega_c)+\sum_p i\frac{\kappa_p}{2}\big)\big] (v-\omega_q) - g^2 }  \\ \times \delta{(u-v)} = t_{21}(v) \delta{(u-v)}.
\label{eq:Smatrix_2port_SinglePhoton_transmission}
\end{multline}
As we will see in Section \ref{sec:TwoPhotonScattering}, these single-photon transport properties will also be useful in analyzing the two-photon transport effects. For later convenience, we also note that the poles of (\ref{eq:Fan_EoM_qubit_lowering_operator_matrix_assump_Fourier_result}) to (\ref{eq:Smatrix_2port_SinglePhoton_transmission}) can be represented as
\begin{multline}
\lambda_{1,\pm} = \frac{\omega_q + \omega_c - i(\kappa_1 + \kappa_2)/2}{2} \\ \pm \sqrt{ \Big\{ \frac{-\omega_q + \omega_c - i ({\kappa_1 + \kappa_2})/2 }{2} \Big\}^2 +g^2}.
\label{eq:Onephoton_poles_lambda1_pm}
\end{multline}

\section{Two-Photon Transport}
\label{sec:TwoPhotonScattering}
Next, we consider the scattering by the cavity-qubit system when two photons are incident from the ports. In similarity to the single-photon case, we can write a general element of the two-photon transport properties as
\begin{multline}
    {}_{b_1,b_2}\langle v_1 v_2^{-}| u_1 u_2^{+}\rangle_{d_1,d_2} = \\ \langle 0|\hat{a}_{\mathrm{out},b_1}(v_1)\hat{a}_{\mathrm{out},b_2}(v_2)\hat{a}^{\dagger}_{\mathrm{in},d_1}(u_1)\hat{a}^{\dagger}_{\mathrm{in},d_2}(u_2)| 0 \rangle.
\label{eq:ScatteringMatrix_TwoPhotons}
\end{multline}
We can simplify the evaluation of this expression by making use of the results from the single-photon transport case. To do this, we can insert a resolution of the identity operator in terms of input states between the two output operators in (\ref{eq:ScatteringMatrix_TwoPhotons}). Due to the nature of the operators involved, the only terms that will remain non-zero in this resolution of the identity are $\int \big[ \,{}_{1}| v^{+} \rangle \langle v^{+} |_{1} + {}_{2}| v^{+} \rangle \langle v^{+} |_{2}\big] \mathrm{d}v$ for the two port case considered here. Doing this, (\ref{eq:ScatteringMatrix_TwoPhotons}) becomes
\begin{multline}
    {}_{b_1,b_2}\langle v_1 v_2^{-}| u_1 u_2^{+}\rangle_{d_1,d_2} \\ =  \int  {}_{b_1}\langle v_1^{-} | v^{+} \rangle_{1} \, {}_{1}\langle v^{+} | \hat{a}_{\mathrm{out},b_2}(v_2)| u_1 u_2 ^{+} \rangle_{d_1,d_2} \, \mathrm{d}v \\ + \int  {}_{b_1}\langle v_1^{-} | v^{+} \rangle_{2} \, {}_{2}\langle v^{+} | \hat{a}_{\mathrm{out},b_2}(v_2)| u_1 u_2 ^{+} \rangle_{d_1,d_2} \, \mathrm{d}v,
\end{multline}
which can be simplified (with a slight abuse of notation) to
\begin{multline}
   {}_{b_1,b_2}\langle v_1 v_2^{-}| u_1 u_2^{+}\rangle_{d_1,d_2} \\ = {}_{b_1}\langle v_1^{-} | v_1^{+} \rangle_{1} \, {}_{1}\langle v_1^{+} | \hat{a}_{\mathrm{out},b_2}(v_2)| u_1 u_2 ^{+} \rangle_{d_1,d_2}  \\ + {}_{b_1}\langle v_1^{-} | v_1^{+} \rangle_{2} \, {}_{2}\langle v_1^{+} | \hat{a}_{\mathrm{out},b_2}(v_2)| u_1 u_2 ^{+} \rangle_{d_1,d_2}.
   \label{eq:ScatteringMatrix_TwoPhotons_ResIden}
\end{multline}
Next, we can leverage the quantum input-output theory boundary condition to rewrite $\hat{a}_{\mathrm{out},b_2}(v_2)$ to get
\begin{multline}
    {}_{b_1,b_2}\langle v_1 v_2^{-}| u_1 u_2^{+}\rangle_{d_1,d_2} \\  = {}_{b_1}\langle v_1^{-} | v_1^{+} \rangle_{1} \bigg[ {}_{1}\langle v_1^{+}|\hat{a}_{\mathrm{in},b_2} (v_2)| u_1 u_2^{+} \rangle_{d_1,d_2}  \\ -i\sqrt{\kappa_{b_2}}\, {}_{1} \langle v_1^{+}|\hat{a}_c(v_2)| u_1 u_2^{+} \rangle_{d_1,d_2}  \bigg] \\ 
    + {}_{b_1}\langle v_1^{-} | v_1^{+} \rangle_{2} \bigg[  {}_{2}\langle v_1^{+}|\hat{a}_{\mathrm{in},b_2} (v_2)| u_1 u_2^{+} \rangle_{d_1,d_2} \\ -i\sqrt{\kappa_{b_2}}\, {}_{2} \langle v_1^{+}|\hat{a}_c(v_2)| u_1 u_2^{+} \rangle_{d_1,d_2} \bigg].
    \label{eq:ScatteringMatrix_TwoPhotons_ResIden_2}
\end{multline}
Recognizing from the orthonormality of the input states that
\begin{multline}
{}_{b_1}\langle v_1^{+} |a_{\mathrm{in},b_2}(v_2) |u_1 u_2^{+} \rangle_{d_1,d_2} \\ = \delta(u_1 - v_1) \delta(u_2 - v_2) \delta_{b_1,d_1} \delta_{b_2,d_2}  
\\ + \delta(u_2 - v_1)  \delta(u_1 - v_2) \delta_{b_1,d_2} \delta_{b_2,d_1},
\label{eq:DoubleDelta_Resulting_Expressions}
\end{multline}
we see that the main quantity that needs to be derived to analyze the two-photon transport is of the general form ${}_{b} \langle v_1^{+}|\hat{a}_c(v_2)| u_1 u_2^{+} \rangle_{d_1,d_2}$.

We can now follow a similar solution procedure as in Section \ref{sec:OnePhotonScattering}, albeit with more tedious algebraic manipulations and results. We start by writing the equations of motion for the relevant matrix elements of (\ref{eq:Fan_EoM_qubit_lowering_operator}) and (\ref{eq:Fan_EoM_c_input_output_QIOT}) to get
\begin{multline}
    \partial_t \, {}_{b}\langle v_1^{+} |\hat{\sigma}_{-}(t)|u_1 u_2^{+} \rangle_{d_1,d_2}  \\ = -i \omega_q \,{}_{b}\langle v_1^{+} |\hat{\sigma}_{-}(t)|u_1 u_2^{+} \rangle_{d_1,d_2}  \\ + ig\, {}_{b}\langle v_1^{+} |\hat{\sigma}_{z}(t) \hat{a}_c(t)|u_1 u_2^{+} \rangle_{d_1,d_2}.
\label{eq:Fan_EoM_qubit_lowering_operator_matrix_TwoPhotonTransport}
\end{multline}
\begin{multline}
    \partial_t \, {}_{b}\langle v_1^{+} |\hat{a}_c(t)|u_1 u_2^{+} \rangle_{d_1,d_2}  \\ = -i\bigg(\omega_c -i \sum_p\frac{\kappa_p}{2} \bigg) {}_{b}\langle v_1^{+} |\hat{a}_c(t)|u_1 u_2^{+} \rangle_{d_1,d_2}  \\
    -\sum_p i\sqrt{\kappa_p} \,{}_{b}\langle v_1^{+} |  \hat{a}_{\mathrm{in},p}(t) |u_1 u_2^{+} \rangle_{d_1,d_2}  \\  - ig \,{}_{b}\langle v_1^{+}|\hat{\sigma}_{-}(t)|u_1 u_2^{+}\rangle_{d_1,d_2},
\label{eq:Fan_EoM_c_input_output_QIOT_matrix_TwoPhotonTransport}
\end{multline}
As with Section \ref{sec:OnePhotonScattering}, we need to simplify the final term in (\ref{eq:Fan_EoM_qubit_lowering_operator_matrix_TwoPhotonTransport}) to proceed. However, in the two-photon case we need a more sophisticated treatment because the qubit can only partially absorb a single photon at a time. To take this into account in a tractable way, we can rewrite $\hat{\sigma}_z  = 2 \hat{\sigma}_{+}\hat{\sigma}_{-} - 1$. We can then insert a resolution of the identity between the qubit operators, with nominally the only terms that produce a non-zero contribution in the final expressions being the $|0\rangle\langle 0|$ portion. Hence, (\ref{eq:Fan_EoM_qubit_lowering_operator_matrix_TwoPhotonTransport}) becomes
\begin{multline}
    \partial_t \, {}_{b}\langle v_1^{+} |\hat{\sigma}_{-}(t)|u_1 u_2^{+} \rangle_{d_1,d_2}  \\ = -i\omega_q \,{}_{b}\langle v_1^{+} |\hat{\sigma}_{-}(t)|u_1 u_2^{+} \rangle_{d_1,d_2}  \\ + ig\, \bigg[ 2\,{}_{b}\langle v_1^{+} |\hat{\sigma}_{+}(t) |0\rangle \langle 0|\hat{\sigma}_{-}(t)\hat{a}_c(t)|u_1 u_2^{+} \rangle_{d_1,d_2} \\
    - {}_{b}\langle v_1^{+} |\hat{a}_c(t)|u_1 u_2^{+} \rangle_{d_1,d_2}\bigg].
\label{eq:Fan_EoM_qubit_lowering_operator_matrix_TwoPhotonTransport_2}
\end{multline}

To have a closed set of equations to solve, we need an equation of motion for $\langle 0|\hat{\sigma}_{-}(t)\hat{a}_c(t)|u_1 u_2^{+} \rangle_{d_1,d_2}$. Evaluating this from the total Hamiltonian of the system, we have
\begin{multline}
    \partial_t \langle 0 |\hat{\sigma}_{-}(t) \hat{a}_c(t)|u_1 u_2^{+} \rangle_{d_1,d_2} \\ = -i\bigg[(\omega_c +\omega_q) -i  \sum_p \frac{\kappa_p}{2} \bigg] \langle 0 |\hat{\sigma}_{-}(t)\hat{a}_c(t)|u_1 u_2^{+} \rangle_{d_1,d_2} \\ - \sum_p i\sqrt{\kappa_p} \langle 0 |\hat{\sigma}_{-}(t) \hat{a}_{\mathrm{in},p}(t) |u_1 u_2^{+} \rangle_{d_1,d_2} \\ - ig \langle 0|\hat{a}_c(t)\hat{a}_c(t)|u_1 u_2^{+}\rangle_{d_1,d_2},
\label{eq:Fan_EoM_qubitc_lowering_operator_matrix_TwoPhotonTransport}        
\end{multline}
which shows that we also need the equation of motion
\begin{multline}
    \partial_t \langle 0 |\hat{a}_c(t)\hat{a}_c(t)|u_1 u_2^{+} \rangle_{d_1,d_2}  \\ = -2i\bigg[\omega_c-  \sum_p i\frac{\kappa_p}{2} \bigg] \langle 0 |\hat{a}_c(t)\hat{a}_c(t)|u_1 u_2^{+} \rangle_{d_1,d_2} \\
    - \sum_p 2i\sqrt{\kappa_p} \langle 0|\hat{a}_c(t)\hat{a}_{\mathrm{in},p}(t)|u_1 u_2\rangle_{d_1,d_2} \\ - 2ig\langle 0 |\hat{\sigma}_{-}(t)\hat{a}_c(t)|u_1 u_2^{+} \rangle_{d_1,d_2}.
\label{eq:Fan_EoM_doublec_input_output_QIOT_matrix_TwoPhotonTransport}
\end{multline}
The set of (\ref{eq:Fan_EoM_c_input_output_QIOT_matrix_TwoPhotonTransport}) to (\ref{eq:Fan_EoM_doublec_input_output_QIOT_matrix_TwoPhotonTransport}) can now be solved together by transforming to the frequency domain in similarity to the process of Section \ref{sec:OnePhotonScattering}. 

The manipulations become somewhat lengthy, so we relegate more of the details to the appendices. The process begins by solving the frequency domain versions of (\ref{eq:Fan_EoM_qubitc_lowering_operator_matrix_TwoPhotonTransport}) and (\ref{eq:Fan_EoM_doublec_input_output_QIOT_matrix_TwoPhotonTransport}) together. The details on evaluating key terms in the Fourier transformation are in Appendix \ref{sec:TwoPhoton_input}, with the final result after solving the two frequency domain equations being 
\begin{multline}
     \langle 0|\hat{\sigma}_{-}(v_2) * \hat{a}_c(v_2)|u_1 u_2^{+} \rangle_{d_1,d_2}   = \\
    \frac{\big[ v_2   -2\omega_c +i  \textstyle \sum_p \kappa_p \big]}{\big[v_2   -2\omega_c +i \textstyle \sum_p \kappa_p \big]  \big[v_2 - \big(\omega_c + \omega_q) + i \textstyle\sum_p \tfrac{\kappa_p}{2}  \big] - 2g^2} \\    
     \times \bigg[ \textstyle \sum_p\sqrt{\kappa_p} \langle 0 |\hat{\sigma}_{-}(v_2) * \hat{a}_{\mathrm{in},p}(v_2) |u_1 u_2^{+} \rangle_{d_1,d_2}      
    \\ + 2g \textstyle \sum_p \sqrt{\kappa_p} \langle 0 |\hat{a}_c(v_2) * \hat{a}_{\mathrm{in},p}(v_2) |u_1 u_2^{+} \rangle_{d_1,d_2}
     \bigg],
     \label{eq:Fan_EoM_qubitc_lowering_operator_matrix_TwoPhotonTransport_Result}
\end{multline}
where 
\begin{multline}
    \langle 0 |\hat{\sigma}_{-}(v_2) * \hat{a}_{\mathrm{in},p}(v_2) |u_1 u_2^{+} \rangle_{d_1,d_2} 
     =   \big[S_{q,d_2}(u_2)  \delta_{p,d_1} \\ + S_{q,d_1}(u_1) \delta_{p,d_2} \big]  \delta \big({ v_2 - (u_1 + u_2) } \big),
	\label{eq:Fan_EoM_qubit_In_matrix_FT}
\end{multline}
\begin{multline}
    \langle 0 |\hat{a}_c(v_2) * \hat{a}_{\mathrm{in},p}(v_2) |u_1 u_2^{+} \rangle_{d_1,d_2} 
     =   \big[S_{c,d_2}(u_2)  \delta_{p,d_1} \\+ S_{c,d_1}(u_1) \delta_{p,d_2} \big]  \delta \big({ v_2 - (u_1 + u_2) } \big).
	\label{eq:Fan_EoM_c_portIn_matrix_FT}
\end{multline}

With (\ref{eq:Fan_EoM_qubitc_lowering_operator_matrix_TwoPhotonTransport_Result}), we can now evaluate the Fourier transform of the key term on the third line of (\ref{eq:Fan_EoM_qubit_lowering_operator_matrix_TwoPhotonTransport_2}) that describes the two-photon interactions between the cavity and qubit. The details of this evaluation are in Appendix \ref{sec:QIOT_TwoPhoton_connected_scattering}, which yields 
\begin{multline}
     {}_{b}\langle v_1^{+} |\hat{\sigma}_{+}(v_2) |0\rangle * \langle 0|\hat{\sigma}_{-}(v_2) * \hat{a}_c(v_2)|u_1 u_2^{+} \rangle_{d_1,d_2}
    \\ =  S^{*}_{q,b}(v_1) \Gamma \big({|u_1 u_2^{+} \rangle_{d_1,d_2}},u_1 + u_2 \big) \\ \times  \delta \big((v_1 + v_2) - (u_1 + u_2) \big), 
\label{eq:Fan_EoM_qubitc_lowering_operator_matrix_TwoPhotonTransport_Alt_part1_FT} 
\end{multline}
where, for a given input state of $|u_1 u_2^{+} \rangle_{d_1,d_2}$ and frequency $\xi$, $\Gamma$ is defined as
\begin{multline}
     \Gamma \big({|u_1 u_2^{+} \rangle_{d_1,d_2}},\xi \big) =  \frac{1}{ \big( \xi - \lambda_{2,+} \big) \big( \xi - \lambda_{2,-}\big)} \\ \times \bigg\{ \big[ \xi   -2\omega_c  + i \sum_p \kappa_p \big] \\ \times  \sum_p\sqrt{\kappa_p} \big[S_{q,d_2}(u_2)  \delta_{p,d_1} + S_{q,d_1}(u_1) \delta_{p,d_2} \big] \\ +  2g  \sum_p \sqrt{\kappa_p} \big[S_{c,d_2}(u_2)  \delta_{p,d_1} + S_{c,d_1}(u_1) \delta_{p,d_2} \big] \bigg\}.
     \label{eq:Fan_EoM_qubitc_lowering_operator_matrix_TwoPhotonTransport_Alt_part1_FT_SimpleNotation_coeff}
\end{multline}
In (\ref{eq:Fan_EoM_qubitc_lowering_operator_matrix_TwoPhotonTransport_Alt_part1_FT_SimpleNotation_coeff}), the poles of the expression are given by
\begin{multline}
\lambda_{2,\pm} = \frac{\omega_q + 3\omega_c - i3(\kappa_1 + \kappa_2)/2}{2} \\ \pm \sqrt{ \Bigg\{ \frac{\omega_q - \omega_c + i ({\kappa_1 + \kappa_2})/2 }{2} \Bigg\}^2 +2g^2},
\label{eq:Twophoton_poles_lambda_pm}
\end{multline}
which will be useful in writing later expressions.


With all the key terms developed, we can now use the frequency domain version of (\ref{eq:Fan_EoM_c_input_output_QIOT_matrix_TwoPhotonTransport}) and (\ref{eq:Fan_EoM_qubit_lowering_operator_matrix_TwoPhotonTransport_2}) to solve for ${}_{b} \langle v_1^{+}|\hat{a}_c(v_2)| u_1 u_2^{+} \rangle_{d_1,d_2}$. The result corresponds to
\begin{multline}
    {}_{b}\langle v_1^{+} |\hat{a}_c(v_2)|u_1 u_2^{+} \rangle_{d_1,d_2} = \\ \frac{1}{(v_2 - \omega_q) \big[ v_2 -\omega_c +i \textstyle \sum_p\tfrac{\kappa_p}{2} \big] -g^2} \\ 
    \times \bigg\{  (v_2 - \omega_q)    \sum_p\sqrt{\kappa_p}\,{}_{b}\langle v_1^{+} |a_{\mathrm{in},p}(v_2) |u_1 u_2^{+} \rangle_{d_1,d_2}        \\
    -  \frac{g^2}{\pi} \,S^{*}_{q,b}(v_1) \Gamma \big({|u_1 u_2^{+} \rangle_{d_1,d_2}},u_1 + u_2 \big)  \\ \times \delta \big((v_1 + v_2) - (u_1 + u_2) \big) \bigg\}.
\label{eq:Fan_EoM_c_input_output_QIOT_matrix_TwoPhotonTransport_Result1_2}    
\end{multline}

This may be substituted into (\ref{eq:ScatteringMatrix_TwoPhotons_ResIden_2}) and the resulting expression simplified for the various possible scattering cases. 

In this work, we will only consider the case where two photons are incident from Port 1 for simplicity. Further, we will assume from here on that the ports of the system are symmetric in the sense that $\kappa_1 = \kappa_2$ as this allows the forthcoming expressions to simplify into a much more compact form. Some of the salient details are shown in Appendix \ref{sec:Smat_general_case} for how to simplify these expressions for the different cases considered. To begin here, we consider the case where both photons are reflected. This yields
\begin{multline}
    {}_{1,1}\langle v_1 v_2^{-}| u_1 u_2^{+}\rangle_{1,1}
    = r_{1}(v_1) r_{1}(v_2) \times \\ [\delta(u_1 - v_1) \delta(u_2 - v_2)  + \delta(u_2 - v_1)  \delta(u_1 - v_2) ]
    \\ + \frac{i g }{\pi} S^{}_{q,1}(v_1) S^{}_{q,1}(v_2) \Gamma \big({|u_1 u_2^{+} \rangle_{1,1}},u_1+u_2 \big) \,\, \times \\ \delta\big((v_1 + v_2) - (u_1 + u_2) \big). 
\label{eq:ScatteringMatrix_TwoPhotons_explicit_main_ScatteringCase1}
\end{multline}
Next, the case where both photons are transmitted to Port 2 yields
\begin{multline}
{}_{2,2}\langle v_1 v_2^{-}| u_1 u_2^{+}\rangle_{1,1}
    = t_{21}(v_1) t_{21}(v_2) \times \\ [\delta(u_1 - v_1) \delta(u_2 - v_2)  + \delta(u_2 - v_1)  \delta(u_1 - v_2) ]
        \\ + \frac{i g }{\pi} S^{}_{q,2}(v_1) S^{}_{q,2}(v_2) \Gamma \big({|u_1 u_2^{+} \rangle_{1,1}},u_1+u_2 \big) \,\, \times \\ \delta\big((v_1 + v_2) - (u_1 + u_2) \big).    
\label{eq:ScatteringMatrix_TwoPhotons_explicit_main_ScatteringCase3}
\end{multline}
The remaining cases where the photons exit separate ports give
\begin{multline}
    {}_{2,1}\langle v_1 v_2^{-}| u_1 u_2^{+}\rangle_{1,1}
    = t_{21}(v_1) r_{1}(v_2) \times \\ [\delta(u_1 - v_1) \delta(u_2 - v_2)  + \delta(u_2 - v_1)  \delta(u_1 - v_2) ]
    \\ + \frac{i g }{\pi} S^{}_{q,2}(v_1) S^{}_{q,1}(v_2) \Gamma \big({|u_1 u_2^{+} \rangle_{1,1}},u_1+u_2 \big) \,\, \times \\ \delta\big((v_1 + v_2) - (u_1 + u_2) \big), 
\label{eq:ScatteringMatrix_TwoPhotons_explicit_main_ScatteringCase2}
\end{multline}
\begin{multline}
    {}_{1,2}\langle v_1 v_2^{-}| u_1 u_2^{+}\rangle_{1,1}
    = r_{1}(v_1) t_{21}(v_2) \times \\ [\delta(u_1 - v_1) \delta(u_2 - v_2)  + \delta(u_2 - v_1)  \delta(u_1 - v_2) ]
        \\ + \frac{i g }{\pi} S^{}_{q,1}(v_1) S^{}_{q,2}(v_2) \Gamma \big({|u_1 u_2^{+} \rangle_{1,1}},u_1+u_2 \big) \,\, \times \\ \delta\big((v_1 + v_2) - (u_1 + u_2) \big).
\label{eq:ScatteringMatrix_TwoPhotons_explicit_main_ScatteringCase4}
\end{multline}
These results will be used in Section \ref{sec:g2} to compute the second-order correlation function that is useful for characterizing the scattering properties of the cavity-qubit system.

\section{Second-Order Correlation Functions}
\label{sec:g2}
As with the end of Section \ref{sec:TwoPhotonScattering}, we specialize our analysis to a system with two symmetric ports so that $\kappa_{1} = \kappa_{2} = \kappa$ to keep the length of the expressions smaller. We will also specifically consider the case where all photons will be incident on the cavity-qubit system from Port $1$. Further, we will use a ``real-space formulation'' like in \cite{rephaeli2012few} to express the wavefunction for different cases of scattering output. Then, using this wavefunction, we can evaluate the second-order correlation $g^{(2)}(\tau)$ for various operating regimes to better interpret the physical effects of the cavity-qubit system.

To utilize this real-space formulation, we need to express the port output states at a single frequency in a position basis. Considering that for our geometry we only consider the fundamental mode of the coaxial regions, we can recognize that the linear dispersion relation gives us a wavenumber of $u/v_p$, where $u$ is the frequency and $v_p$ is the phase velocity. We can then utilize a spatial Fourier transform to relate frequency and position basis operators. Under such an approach, we can denote an output creation operator in Port $b$ at frequency $u$ as
\begin{align}
    \hat{a}^\dagger_{\mathrm{out},b}(u) = \frac{1}{\sqrt{2 \pi}}\int  \hat{a}^\dagger_{b}(x) e^{i xu/v_g} \mathrm{d} x,
\label{eq:RealSpace_port_FT}    
\end{align}
where the position operators satisfy the commutation relationship 
\begin{align}
    [\hat{a}_{p}(x), \hat{a}^{\dagger}_{b}(x^{\prime})] = \delta_{pb}\delta(x - x^{\prime}).
    \label{eq:pos_commutation}
\end{align}
Then, when considering the representation of two photons, this can become
\begin{multline}
    \hat{a}^\dagger_{\mathrm{out},b_1}(u_1) \hat{a}^\dagger_{\mathrm{out},b_2}(u_2)    =  \iint   P(u_1,u_2; x_1,x_2) \\ \times \frac{1}{\sqrt{2}}\hat{a}^\dagger_{b_1}(x_1)  \hat{a}^\dagger_{b_2}(x_2) \mathrm{d} x_1 \mathrm{d} x_2,
\label{eq:RealSpace_port_FT2}    
\end{multline}
where 
\begin{multline}
P(u_1,u_2;x_1,x_2) = \frac{1}{\sqrt{2} 2 \pi} \big[e^{i x_1 u_1/v_g}  e^{i x_2 u_2/v_g} \\  + e^{i x_1 u_2/v_g}  e^{i x_2 u_1/v_g} \big]   
\end{multline}
and the various normalization factors are needed to ensure the overall quantum state is normalized as well.



Now, we can represent the different components of the output state using the results derived in Section \ref{sec:TwoPhotonScattering}. Applying a resolution of the identity in the output state basis applied to the input state gives us terms of the form
\begin{multline}
|\phi\rangle_{b_1,b_2} = \frac{1}{2} \iint_{-\infty}^{\infty}  \mathrm{d}v_1 \mathrm{d}v_2 \, {}_{b_1,b_2}\langle v_1 v_2^{-}| u_1 u_2 ^{+} \rangle_{1,1}  \\ \times |v_1 v_2 ^{-} \rangle_{b_1,b_2}   ,
\label{eq:general_output_state_FT_definition}
\end{multline}
where the total output state is
\begin{align}
    |\phi\rangle = \sum_{b_1, b_2=1}^2 |\phi\rangle_{b_1,b_2}.
\end{align}
Evaluating (\ref{eq:general_output_state_FT_definition}) depends on which output state component one is interested in inspecting, so we demonstrate the process here for the case where both photons are reflected. Substituting (\ref{eq:ScatteringMatrix_TwoPhotons_explicit_main_ScatteringCase1}) and (\ref{eq:RealSpace_port_FT2}) into the above and evaluating the trivial integrals involving Dirac deltas yields
\begin{multline}
|\phi\rangle_{1,1} =  \iint \mathrm{d} x_1 \mathrm{d} x_2 \Big\{  P(u_1,u_2; x_1,x_2) \\ r_{1}(u_1) r_{1}(u_2)
     \Big\}  \frac{1}{\sqrt{2}} \hat{a}^\dagger_{1}(x_1)  \hat{a}^\dagger_{1}(x_2)  |0\rangle
\\ +  \iint \mathrm{d} x_1 \mathrm{d} x_2    \Big\{ \frac{1}{2} \iint_{-\infty}^{\infty} \mathrm{d}v_1 \mathrm{d}v_2  P(v_1,v_2; x_1,x_2)   \\ 
    \times \frac{i g }{\pi} S^{}_{q,1}(v_1) S^{}_{q,1}(v_2) \Gamma \big({|u_1 u_2^{+} \rangle_{1,1}},u_1+u_2 \big) \\ \times \delta\big((v_1 + v_2) - (u_1 + u_2) \big) \Big\} \frac{1}{\sqrt{2}}\hat{a}^\dagger_{1}(x_1)  \hat{a}^\dagger_{1}(x_2)|0\rangle.
\label{eq:general_output_state_FT_rr2}
\end{multline}


Evaluating the remaining frequency integrals requires further care due to the poles of the overall expression. The corresponding contour integrals are detailed in Appendix \ref{sec:Contour_Integral}, where it is convenient to write the expressions in terms of the total incoming and outgoing frequencies as $E_i = u_1 + u_2$ and $E_o = v_1 + v_2$. Further changes of variables for the spatial coordinates are also convenient in the form of $x_m = x_1 - x_2$ and $x_c = (x_1+x_2)/2$. Then, we can find that (\ref{eq:general_output_state_FT_rr2}) after the manipulations in Appendix \ref{sec:Contour_Integral} becomes
\begin{multline}
|\phi\rangle_{1,1} = \iint \mathrm{d}x_1 \mathrm{d}x_2 
\bigg[ r_{1}(u_1) r_{1}(u_2) P(u_1,u_2;x_1,x_2)  \\ + H(x_1,x_2) \bigg]
\frac{1}{\sqrt{2}} \hat{a}^{\dagger}_{1}(x_1) \hat{a}^{\dagger}_{1}(x_2)|0\rangle,
\label{eq:output_state_rr}
\end{multline}
where
\begin{multline}
H(x_1,x_2) = \frac{i g }{\pi} \Gamma \big({|u_1 u_2^{+} \rangle_{1,1}},E_i \big)     \\  \times\frac{i g^2 \kappa }{\sqrt{2} }      \frac{e^{i E_i x_c/v_g}}{( \lambda_{1,+}  - \lambda_{1,-}) ({E_i} - \lambda_{1,+} - \lambda_{1,-})}  \\ \times \Bigg[ \frac{ e^{i ({E_i} - 2\lambda_{1,-}) \tfrac{|x_m|}{2v_g}} }{({E_i} - 2\lambda_{1,-})   }  
 - \frac{ e^{i ({E_i} - 2\lambda_{1,+}) \tfrac{|x_m|}{2v_g} }}{({E_i} - 2\lambda_{1,+})  }  \Bigg].
\label{eq:H_general_final_result}
\end{multline}

The above procedure can also be applied to the remaining different output scattering state components. This yields
\begin{multline}
|\phi\rangle_{2,2} = \iint \mathrm{d}x_1 \mathrm{d}x_2 
 \bigg[ t_{21}(u_1) t_{21}(u_2) P(u_1,u_2;x_1,x_2) \\ + H(x_1,x_2)
\bigg] 
\frac{1}{\sqrt{2}} \hat{a}^{\dagger}_{2}(x_1) \hat{a}^{\dagger}_{2}(x_2) |0\rangle,
\label{eq:output_state_tt}
\end{multline}
\begin{multline}
|\phi\rangle_{2,1} = \iint \mathrm{d}x_1 \mathrm{d}x_2 
 \bigg[ \frac{1}{2} \big(t_{21}(u_1) r_{1}(u_2) + r_1(u_1) t_{21}(u_2) \big)  \\ \times P(u_1, u_2;x_1,x_2)  + H(x_1,x_2)
\bigg] 
\frac{1}{\sqrt{2}} \hat{a}^{\dagger}_{2}(x_1) \hat{a}^{\dagger}_{1}(x_2) |0\rangle,
\label{eq:output_state_tr}
\end{multline}
\begin{multline}
|\phi\rangle_{1,2} = \iint \mathrm{d}x_1 \mathrm{d}x_2 
 \bigg[ \frac{1}{2} \big( r_{1}(u_1) t_{21}(u_2) + t_{21}(u_1) r_1(u_2) \big) \\ \times P(u_1,u_2;x_1,x_2)  + H(x_1,x_2) 
\bigg] 
\frac{1}{\sqrt{2}} \hat{a}^{\dagger}_{1}(x_1) \hat{a}^{\dagger}_{2}(x_2) |0\rangle.
\label{eq:output_state_rt}
\end{multline}


Using the above expressions, the photon statistics of the two photons are characterized by the second-order correlation function $g^{(2)}(\tau)$. Various definitions for computing $g^{(2)}(\tau)$ exist, so we use the one most advantageous for the formalism used here. The general expression is \cite{shi2011two, rephaeli2012few, xu2016fano}
\begin{align}
    g^{(2)}_{b_1,b_2}(\tau) = \frac{G^{(2)}_{b_1,b_2}(\tau)}{G^{(2)}_{b_1,b_2}(\tau \rightarrow \infty)},
\end{align}
where we have introduced the subscripts $b_1$ and $b_2$ to the various second-order correlation functions to aid in differentiating the different cases of interest for this system and $G^{(2)}_{b_1,b_2}(\tau)$ is defined as
\begin{multline}
G^{(2)}_{b_1,b_2}(\tau) = \langle \phi| \hat{a}_{b_1}^{\dagger}(x) \hat{a}_{b_2}^{\dagger}(x + v_g\tau) \times \\
\hat{a}_{b_2}(x + v_g\tau) \hat{a}_{b_1}(x) |\phi \rangle.
\label{eq:G2Tau_Xu_GeneralExpression}
\end{multline}

The computation of $G^{(2)}_{b_1,b_2}(\tau)$ can be accomplished easily by leveraging the commutation relation of the creation and annihilation operators given in (\ref{eq:pos_commutation}). Doing this, we see that for the case where two photons are reflected back to Port $1$ we have
\begin{multline}
G^{(2)}_{1,1}(\tau)  = 2|r_1(u_1)r_1(u_2)P(u_1,u_2; x, x+v_g\tau) \\ + H(x, x+v_g\tau)|^{2}.
\label{eq:G2Tau_Xu_GeneralExpression_appendix_rr}
\end{multline}
Similarly, the case where the two photons are transmitted to Port $2$ gives
\begin{multline}
G^{(2)}_{2,2}(\tau)  = 2|t_{21}(u_1)t_{21}(u_2)P(u_1,u_2; x, x+v_g\tau) \\ + H(x, x+v_g\tau)|^{2}.
\label{eq:G2Tau_Xu_GeneralExpression_tt}
\end{multline}
Finally, when the two photons exit through separate ports the result is
\begin{multline}
G^{(2)}_{2,1}(\tau)  = 2 \bigg| \frac{1}{2} \big( r_1(u_1)t_{21}(u_2) + t_{21}(u_1) r_1(u_2) \big) \\ \times P(u_1,u_2;x,x+v_g\tau)  + H(x, x+ v_g \tau) \bigg|^{2}.
\label{eq:G2Tau_Xu_GeneralExpression_rt}
\end{multline}

\section{Result}
\label{sec:result}
In this section, we evaluate the transport characteristics of single- and two-photon scattering through a cavity-qubit system using the analytical solutions developed throughout this work. We begin by considering the system operated in the good-cavity regime of cavity QED where $g > \kappa$ in Section \ref{subsec:gcl}. Following this, we look at the same transport properties in the bad-cavity regime where $g < \kappa$ in Section \ref{subsec:bcl}.

In all cases, we consider a system like that shown in Fig. \ref{fig:cavity_qubit_port_geometries} where there is a rectangular waveguide cavity coupled to by two coaxial ports. The size of the waveguide cavity is $22.86 \times 10.16 \times 40 \, \mathrm{mm}^3$, which leads to a fundamental mode frequency of $\omega_c/2\pi = 7.55 \, \mathrm{GHz}$. The dimensions of the coaxial regions are $r_\mathrm{in} = 0.05\, \mathrm{mm}$ and $r_\mathrm{out} = 2.5\,\mathrm{mm}$ with material properties of $\epsilon_r = 22.04$ and $\mu_r = 1$. Further, the inner conductor of the coaxial ports does not protrude into the cavity region to keep the coupling strength at a reasonable level to allow studying both the good- and bad-cavity regimes. These two coaxial regions are further placed symmetrically about the center of the cavity and are $20\,\mathrm{mm}$ apart from each other to achieve the same magnitude of the cavity-port coupling; more specifically, $g_{p_1} = -g_{p_2}$. Leveraging the microwave cavity perturbation theory discussed in \cite{Moon2024Analytical}, these $g_p$'s can be evaluated and found to provide a cavity decay rate through each port of $\kappa/2\pi = 421.5\,\mathrm{kHz}$.

The transmon qubit inside the cavity is modeled using a small wire dipole of length $1\,\mathrm{mm}$ with a load capacitance of $50.34\,\mathrm{fF}$ and a Josephson junction with parameters set so that $\omega_q/2\pi = 7.55\,\mathrm{GHz}$. This is selected to match $\omega_c$ to demonstrate more clearly the interesting physics of the cavity-qubit system due to the quantum interactions of the qubit and cavity fields. The final parameter needed to evaluate the analytical solutions is the coupling rate between the fundamental mode of the cavity and the transmon qubit, $g$, given in (\ref{eq:coupling_strength_g}). We can again use the techniques from \cite{Moon2024Analytical} to evaluate this, which evaluates the necessary integral of the electric field mode using the receiving characteristics of the dipole antenna. Critically, by keeping the dipole antenna small the fundamental field mode as seen by the dipole can be approximated as appearing like a plane wave, which allows free-space receiving characteristics of the dipole antenna to be used in evaluating the needed properties. In order to achieve $g$'s in the good- and bad-cavity regimes, we reposition the dipole antenna throughout the cavity and also tilt its orientation relative to the fundamental cavity field mode.

\subsection{Good-Cavity Operating Regime}
\label{subsec:gcl}
For the good-cavity operating regime, we maximize the coupling between the qubit and cavity field mode by placing the dipole at the center of the cavity and align its arms parallel to the fundamental cavity field mode electric field. This gives $g = 15.9\,\mathrm{MHz}$. We now analyze the photon transport properties for the single- and two-photon cases when the photon frequencies are near to the qubit and cavity resonant frequencies of $7.55 \, \mathrm{GHz}$. 

\begin{figure}[t!]
    \centering
    \includegraphics[width=0.91\linewidth]{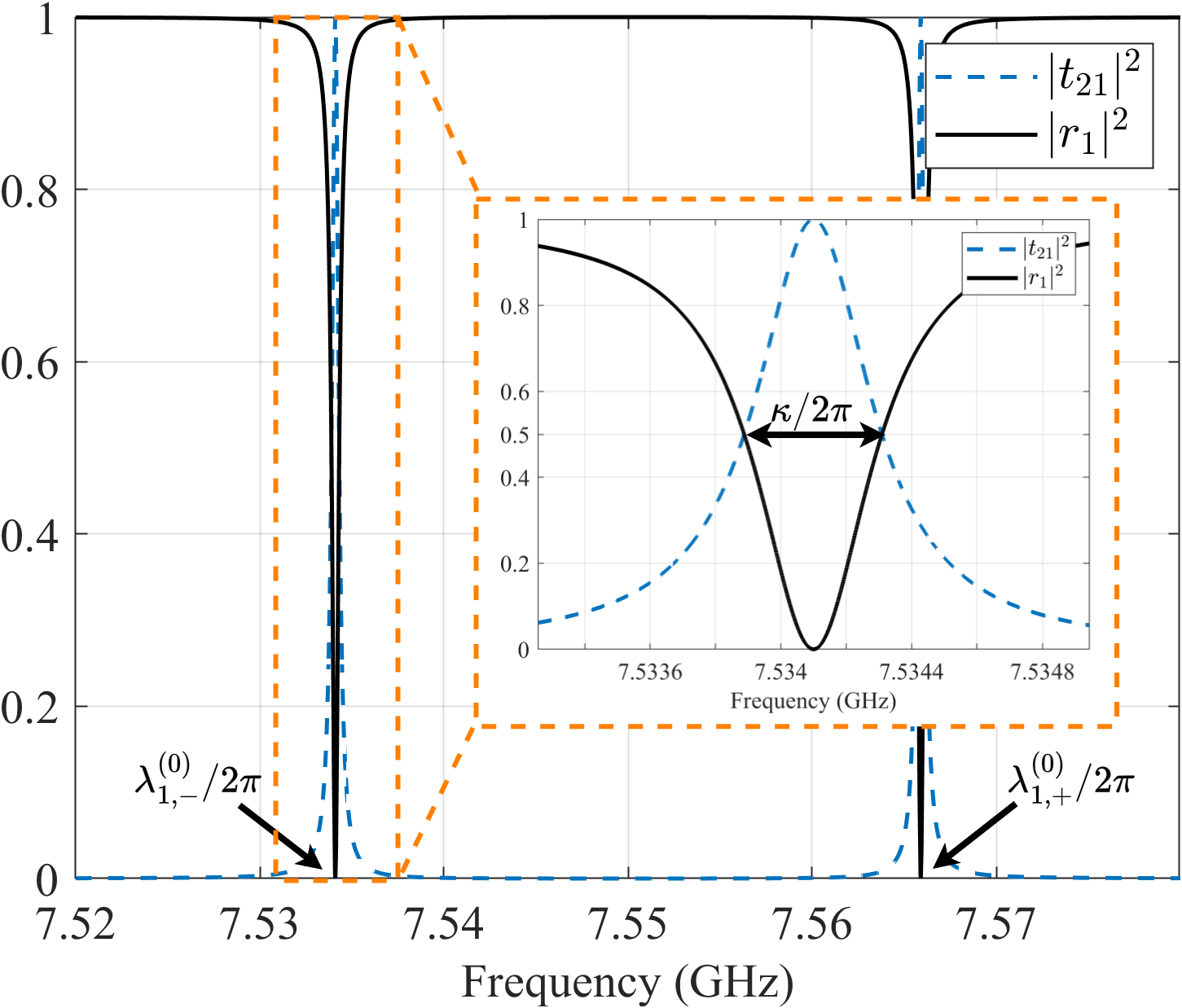} 
    \caption{The single-photon transmission and reflection coefficients for the cavity-qubit system when in the good-cavity regime.}
    \label{fig:SinglePhotonTransport_tr_SPT_Good}
\end{figure}

We begin by showing the single photon transmission and reflection characteristics in Fig. \ref{fig:SinglePhotonTransport_tr_SPT_Good}. The plot shows two clearly separated resonances where the transmission through the cavity peaks. These two transmission peaks correspond to the frequencies of the dressed cavity-qubit system due to the vacuum Rabi splitting of the system. The separation between the two peaks is given by $2g$, which matches the dressed resonant frequencies of the cavity-qubit system denoted by $\lambda_{1,\pm}^{(0)} = \omega_c \pm g$. From the inset image in Fig. \ref{fig:SinglePhotonTransport_tr_SPT_Good}, it is also clear that the full-width half-maximum of the cavity-qubit resonances matches $\kappa$.

\begin{figure}[t!]
    \centering
    \includegraphics[width=0.95\linewidth]{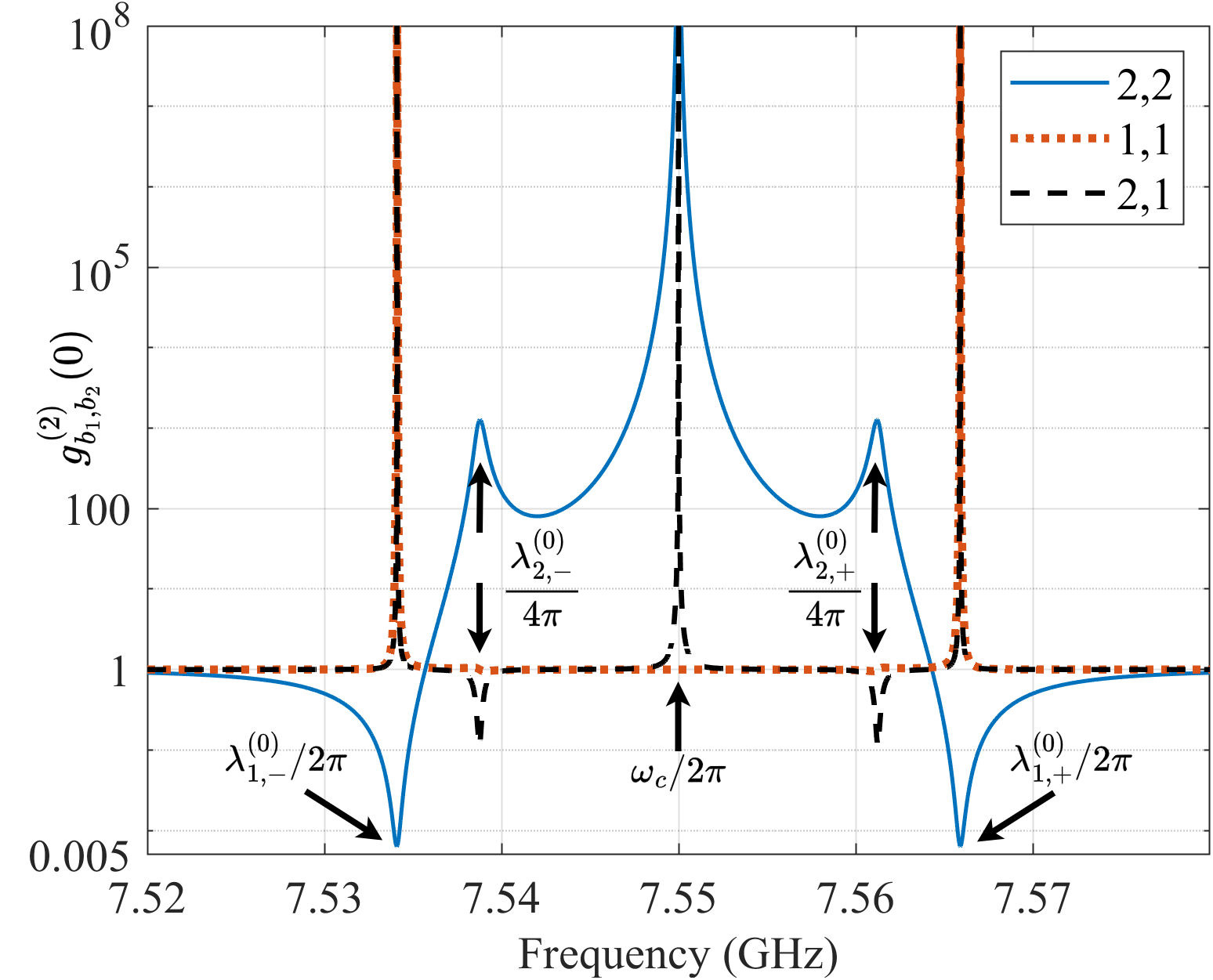}  
    \caption{The second-order correlation function $g^{(2)}_{b_1,b_2}(0)$ in the good-cavity regime. The legend denotes the values of $b_1,b_2$ which focus on the correlation function for the different output scattering cases.}
    \label{fig:TwoPhotonTransport_g20_GCL}
\end{figure}

Next, we analyze the two-photon transport properties for this system. We consider the case where the two photons are both incident from Port 1 and have identical frequencies. The various second-order correlation functions for the output photons are shown in Fig. \ref{fig:TwoPhotonTransport_g20_GCL}. From this, we see that there are a few critical locations for the transport properties. The first occurs near the single-photon dressed resonant frequencies of the cavity-qubit system denoted by $\lambda_{1,\pm}^{(0)}$. In particular, we see that the behavior changes dramatically in comparison to the single-photon case. In the single-photon case, the transmission is strongly facilitated through the cavity-qubit system at this frequency. However, in the two-photon case we no longer have the incident two-photon energy matching the dressed eigenstates of the cavity-qubit system and so we have strong suppression of the two photons being able to simultaneously transmit through the cavity-qubit system. 

Another interesting operating point occurs at the two-photon dressed eigenstates of the cavity-qubit system, which correspond to $\lambda_{2,\pm}^{(0)} = \omega_c \pm \sqrt{2}g$. At this frequency, the incident two-photon energy matches a dressed eigenstate of the cavity-qubit system and we see that both photons can be allowed to transmit through the cavity-qubit system simultaneously. Again, this behavior is dramatically different from the single-photon transport properties at this frequency where the single-photon transmission is essentially completely suppressed. We also see a pronounced ability for the photons to transmit through the cavity-qubit system at the bare resonant frequency $\omega_c$ where in the single-photon case essentially only reflection occurs.

\subsection{Bad-Cavity Operating Regime}
\label{subsec:bcl}
For the bad-cavity operating regime, we need to lower the coupling between the qubit and cavity field mode. We accomplish this here by moving the dipole to the point $(-6.43 \, \mathrm{mm}, 0 \, \mathrm{mm}, -15 \, \mathrm{mm})$ inside the cavity, where the origin of the cavity coordinate system is located at the center of the cavity. We further rotate the dipole arms so that they make an $86^\circ$ angle with respect to the cavity fundamental field mode orientation. Accounting for these changes, we have that $g = 269.3\,\mathrm{kHz}$. We keep all the other basic settings regarding the incident photon frequencies the same as in Section \ref{subsec:gcl}.

\begin{figure}[t!]
    \centering
    \includegraphics[width=0.91\linewidth]{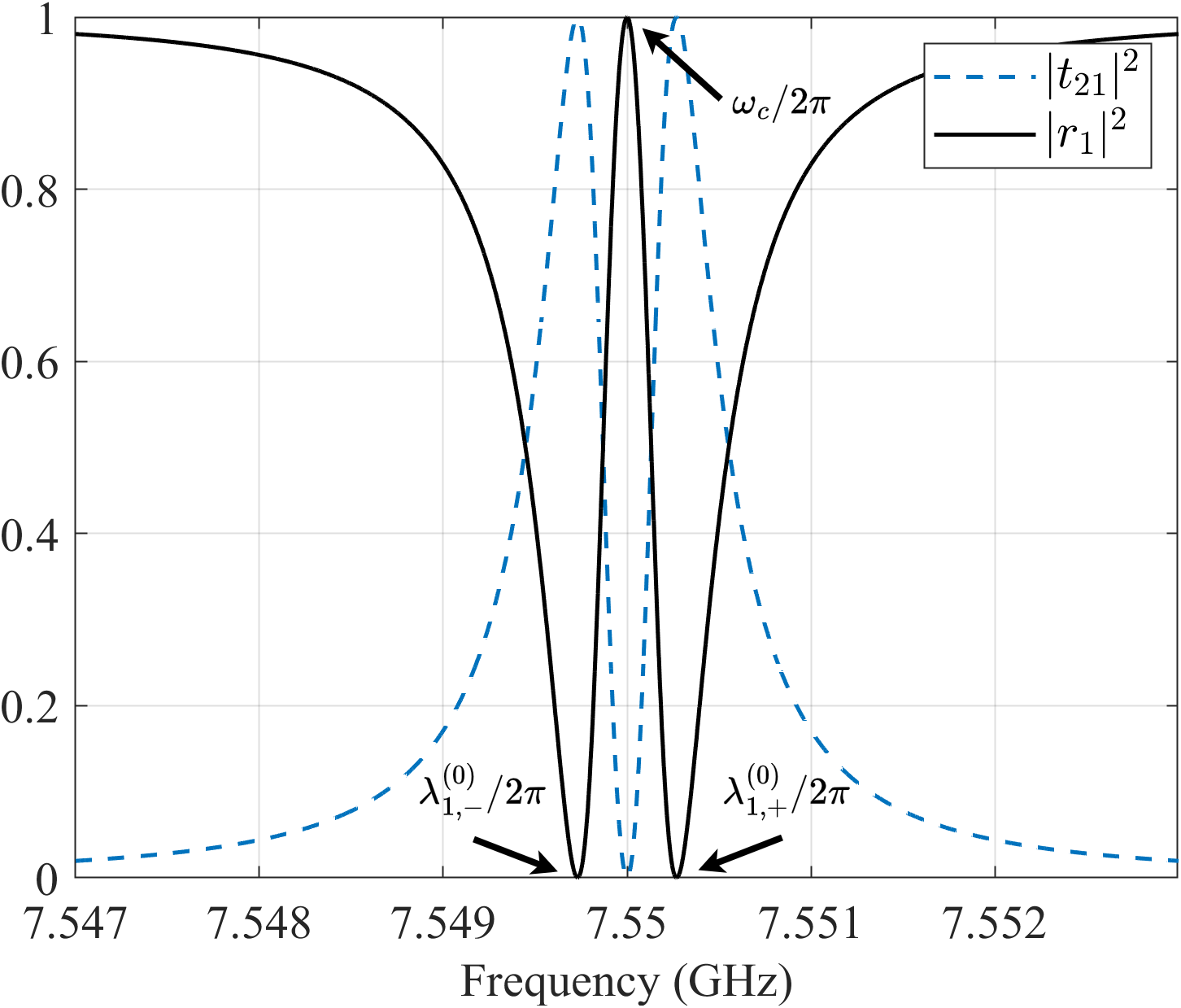} 
    \caption{The single-photon transmission and reflection coefficients for the cavity-qubit system when in the bad-cavity regime.}
    \label{fig:SinglePhotonTransport_tr_result_SPT_Bad}
\end{figure}

We first show the single-photon transmission and reflection coefficients in Fig. \ref{fig:SinglePhotonTransport_tr_result_SPT_Bad}. We see that in the bad-cavity regime there is a new interesting feature at the cavity fundamental frequency that is more visible due to the minimal shifting of the single-photon cavity-qubit dressed eigenstates when $g < \kappa$. This is an analog of the dipole-induced transparency effect, where in this case due to the direct coupling of the ports to the cavity we have an abrupt transition to reflective behavior from the cavity-qubit system.

\begin{figure}[t!]
    \centering
    \includegraphics[width=0.95\linewidth]{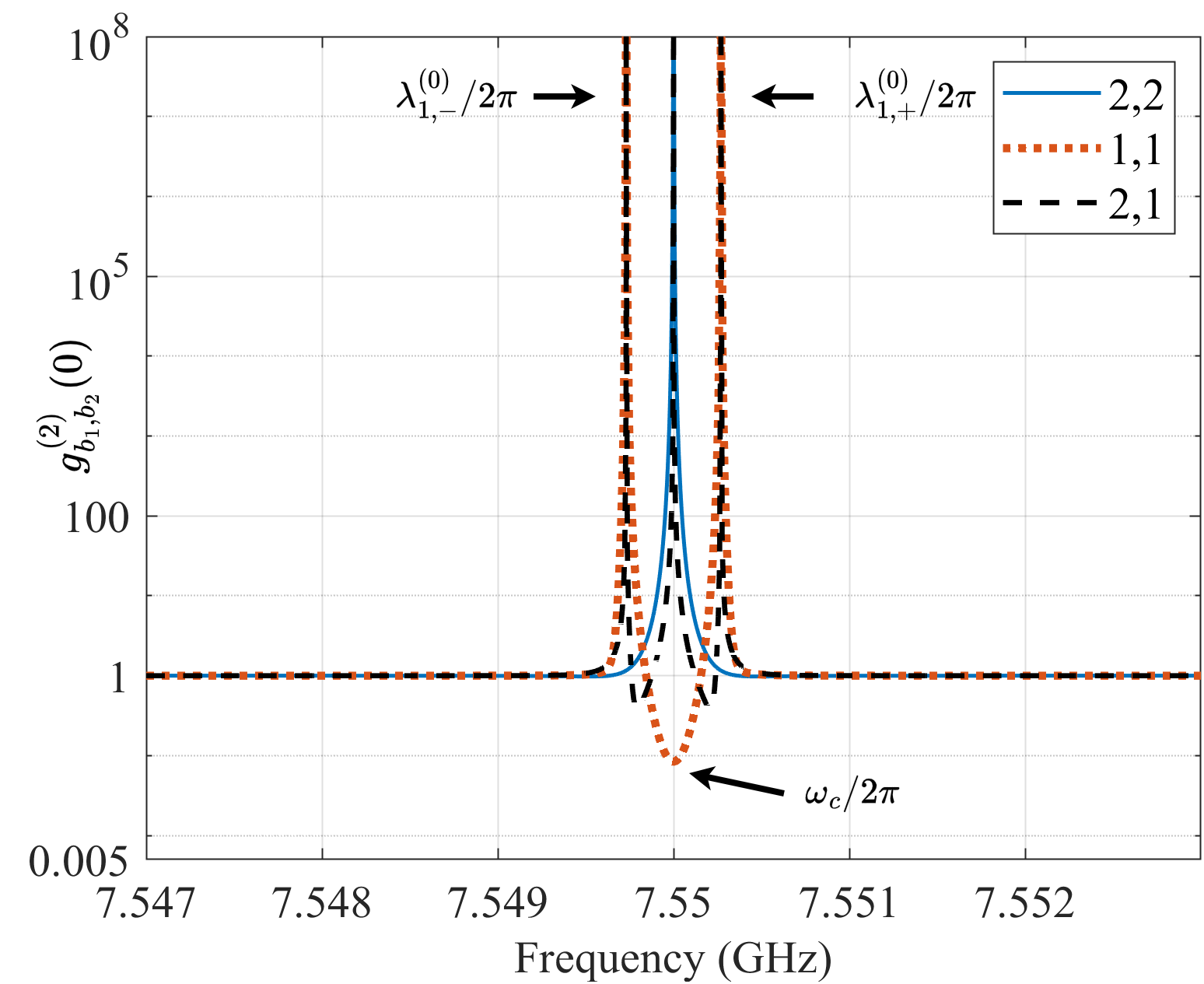} 
    \caption{The second-order correlation function $g^{(2)}_{b_1,b_2}(0)$ in the bad-cavity regime. The legend denotes the values of $b_1,b_2$ which focus on the correlation function for the different output scattering cases.}
    \label{fig:TwoPhotonTransport_g20_BCL}
\end{figure}

Moving now to the two-photon transport effects, we show the second-order correlation functions in Fig. \ref{fig:TwoPhotonTransport_g20_BCL}. As in the case of the good-cavity regime, we see a dramatic change to the properties between single- and two-photon cases. However, the main change now happens near the fundamental cavity resonance frequency. In this case, we see that the single-photon behavior is purely reflective but that in the two-photon case transmission becomes possible.

\section{Conclusion}
\label{sec:conclusion}
In this work, we have developed an analytical quantum full-wave solution to characterize the single- and two-photon transport properties through a cavity-qubit system. The full-wave aspects of the analysis were evaluated using classical electromagnetic theory techniques from microwave and antenna engineering following the process originally established in \cite{Moon2024Analytical}. The analytical solution for the photon transport aspects were developed leveraging a quantum input-output theory formalism commonly used in quantum optics. Numerical results were generated for a cavity-qubit system that is interfaced with through two coaxial transmission lines with device parameters selected to be within the regions of validity for the classical electromagnetic theory approximations needed in the analysis. This analysis was done for the cavity-qubit system in the good- and bad-cavity regimes of cavity quantum electrodynamics, providing a range of unique characteristic features in the single- and two-photon transport properties that can be used to help validate future quantum full-wave numerical methods. 

In the future, our approach can be extended to consider additional cavity-qubit systems (e.g., cylindrical cavities). Other useful extensions would include accounting for the effects of more energy levels in the characterization of the transmon qubit, as these can be important for accurately analyzing different experimental systems. More sophisticated analytical electromagnetic theory techniques could also be explored to expand the range of device parameters where analytical solutions can be found reliably.

\appendix
\renewcommand{\thesection}{\Alph{section}}
\section{Quantum Input-Output Theory Commutation Relations}
\label{sec:QIOT_operator_commutation}
Here, we derive the commutation relations of the input and output field mode operators of a port in the time and frequency domains. To derive the time domain commutation relationship, we substitute the inverse Fourier transform definition for each of the input field mode operators. Then, the commutation relation is expressed as 
\begin{multline}
[\hat{a}_{\mathrm{in},p}(t), \hat{a}^{\dagger}_{\mathrm{in},b}(t^{\prime})] 
 \\ = \frac{1}{2 \pi}\int_{-\infty}^{\infty}\int_{-\infty}^{\infty}  e^{-iu(t-t_0)}\hat{a}_{p}(t_0, u)  \times  \\ e^{iv(t^{\prime}-t_0)}\hat{a}^{\dagger}_{b}(t_0,v) \mathrm{d}u \mathrm{d}v  \\ 
-\frac{1}{2 \pi}\int_{-\infty}^{\infty}\int_{-\infty}^{\infty}  e^{iv(t^{\prime}-t_0)}\hat{a}^{\dagger}_{b}(t_0,v)  \times \\ e^{-iu(t-t_0)}\hat{a}_{p}(t_0,u)  \, \mathrm{d}u \mathrm{d}v.
\label{eq:input_commutator_relationship1}
\end{multline}
Consolidating terms, this can be written as
\begin{multline}
[\hat{a}_{\mathrm{in},p}(t), \hat{a}^{\dagger}_{\mathrm{in},b}(t^{\prime})] 
\\ = \frac{1}{2 \pi}\int_{-\infty}^{\infty}\int_{-\infty}^{\infty}
{e^{i(u-v)t_0}} 
e^{-iut}e^{iv t^{\prime}} \times \\
[\hat{a}_{p}(t_0,u), 
\hat{a}^{\dagger}_{b}(t_0,v)] \mathrm{d}u \mathrm{d}v.
\label{eq:input_commutator_relationship2}
\end{multline}
Using the commutation relation of the port operators given in (\ref{eq:port_field_commutation}), the corresponding Dirac delta that is introduced allows us to evaluate one integral to reduce (\ref{eq:input_commutator_relationship2}) to
\begin{align}
[\hat{a}_{\mathrm{in},p}(t), \hat{a}^{\dagger}_{\mathrm{in},b}(t^{\prime})] 
= \frac{1}{2 \pi}\int_{-\infty}^{\infty} 
e^{-iu (t-t^{\prime}) } \delta_{p,b} \mathrm{d}u.
\label{eq:input_commutator_relationship3}
\end{align}
The remaining integral can be evaluated easily from basic Fourier theory, resulting in
\begin{align}
[\hat{a}_{\mathrm{in},p}(t), \hat{a}^{\dagger}_{\mathrm{in},b}(t^{\prime})] = \delta(t-t^{\prime})\delta_{p,b}.
\label{eq:input_commutator_relationship_final}
\end{align}
Using the same procedure, the output field operator yields the same result of
\begin{align}
[\hat{a}_{\mathrm{out},p}(t), \hat{a}^{\dagger}_{\mathrm{out},b}(t^{\prime})] 
= \delta(t-t^{\prime})\delta_{p,b}.
\label{eq:input_output_commutator_relationship_final2}
\end{align}

Now, we show the derivation of the frequency domain commutator. Similarly as before, we repeat substituting the Fourier transform definition for the input field operator of (\ref{eq:ain_freq_FT}) to express each of the frequency domain operators. As a result, the commutator can be compactly expressed as
\begin{multline}
[\hat{a}_{\mathrm{in},p}(u), \hat{a}^{\dagger}_{\mathrm{in},b}(v)] 
\\ = \frac{1}{2 \pi}\int_{-\infty}^{\infty}\int_{-\infty}^{\infty} 
e^{iut}e^{-iv t^{\prime}}  
[\hat{a}_{\mathrm{in},p}(t), 
\hat{a}^{\dagger}_{\mathrm{in},b}(t^{\prime})] \mathrm{d}t \mathrm{d}t^{\prime}.
\label{eq:input_commutator_relationship_freq2}
\end{multline}
Using the result from (\ref{eq:input_commutator_relationship_final}) for the time domain commutator, we see that a similar set of steps as above allows us to determine that
\begin{align}
[\hat{a}_{\mathrm{in},p}(u), \hat{a}^{\dagger}_{\mathrm{in},b}(v)] 
 = \delta(u - v) 
\delta_{p,b}.
\label{eq:input_commutator_relationship_freq_final}
\end{align}
By repeating the same approach, the commutation relation of the frequency domain output field operator is evaluated to 
\begin{align}
[\hat{a}_{\mathrm{out},p}(u), \hat{a}^{\dagger}_{\mathrm{out},b}(v)] 
= \delta(u - v) 
\delta_{p,b}.
\label{eq:input_output_commutator_relationship_freq_final}
\end{align}

\section{Qubit and Cavity Response to Photons}
\label{sec:TwoPhoton_input}
In this appendix, we evaluate the Fourier transform of expressions $\langle 0 |\hat{\sigma}_{-}(t) \hat{a}_{\mathrm{in},p}(t) |u_1 u_2^{+} \rangle_{d_1,d_2}$ and $\langle 0 |\hat{a}_c(t)  \hat{a}_{\mathrm{in},p}(t) |u_1 u_2^{+} \rangle_{d_1,d_2}$ that arise as terms in the two-photon transport equations of motion (\ref{eq:Fan_EoM_qubitc_lowering_operator_matrix_TwoPhotonTransport}) and (\ref{eq:Fan_EoM_doublec_input_output_QIOT_matrix_TwoPhotonTransport}). This process can be simplified by first expressing $\hat{a}_{\mathrm{in},p}(t)$ through its inverse Fourier transform definition. After substitution into the expression involving $\hat{\sigma}_{-}(t)$, we have 
\begin{multline}
     \langle 0 |\hat{\sigma}_{-}(t) \hat{a}_{\mathrm{in},p}(t) |u_1 u_2^{+} \rangle_{d_1,d_2} \\ = \frac{1}{\sqrt{2 \pi}}\int_{-\infty}^{\infty}   \langle 0 |\hat{\sigma}_{-}(t)\hat{a}_{\mathrm{in},p}(v) |u_1 u_2^{+} \rangle_{d_1,d_2} \,\,e^{-ivt} \mathrm{d}v.
\label{eq:Fan_EoM_qubit_In_matrix_time1}    
\end{multline}
Recognizing that $\langle 0| \hat{a}_{\mathrm{in},p}(v) = {}_{p}\langle v^{+}|,$ the corresponding bosonic symmetry of the system then gives us 
\begin{multline}
     \langle 0 |\hat{\sigma}_{-}(t) \hat{a}_{\mathrm{in},p}(t) |u_1 u_2^{+} \rangle_{d_1,d_2} \\ = \frac{1}{\sqrt{2 \pi}}\int_{-\infty}^{\infty} \,\, {}_{p}\langle v^{+}|u_1^{+} \rangle_{d_1} \langle 0 |\hat{\sigma}_{-}(t)|u_2^{+} \rangle_{d_2} \,\,e^{-ivt} \mathrm{d} v 
    \\ + \frac{1}{\sqrt{2 \pi}}\int_{-\infty}^{\infty}  \,\, {}_{p}\langle v^{+}|u_2^{+} \rangle_{d_2} \langle 0 |\hat{\sigma}_{-}(t)|u_1^{+} \rangle_{d_1} \,\,e^{-ivt} \mathrm{d}v.
\end{multline}
We can simplify this using the orthonormality of the input port states to get
\begin{multline}
     \langle 0 |\hat{\sigma}_{-}(t) \hat{a}_{\mathrm{in},p}(t) |u_1 u_2^{+} \rangle_{d_1,d_2}  = \frac{1}{\sqrt{2\pi}} \langle 0 |\hat{\sigma}_{-}(t) |u_2^{+} \rangle_{d_2} \, \delta_{p,d_1} \, e^{-i u_1 t} \\ + \frac{1}{\sqrt{2\pi}} \langle 0 |\hat{\sigma}_{-}(t) |u_1^{+} \rangle_{d_1} \, \delta_{p,d_2} \, e^{-i u_2 t}.
     \label{eq:sm_ain_2}
\end{multline}
From the single-photon transport analysis in Section \ref{sec:OnePhotonScattering}, we can recognize that the inverse Fourier transform of (\ref{eq:Fan_EoM_qubit_lowering_operator_matrix_assump_Fourier_result}) gives us that
\begin{align}
    \langle 0 |\hat{\sigma}_{-}(t)|u^{+} \rangle_{p} = \frac{1}{\sqrt{2 \pi}} S_{q,p}(u) e^{-iut}.
\end{align}
Using this in (\ref{eq:sm_ain_2}), we have that
\begin{multline}
\langle 0 |\hat{\sigma}_{-}(t) \hat{a}_{\mathrm{in},p}(t) |u_1 u_2^{+} \rangle_{d_1,d_2}    \\ = \frac{1}{2\pi}  \big[ S_{q,d_2}(u_2)  \delta_{p,d_1} + S_{q,d_1}(u_1) \delta_{p,d_2} \big]  e^{-i (u_1 + u_2) t}.
\label{eq:Fan_EoM_qubit_In_matrix_time2}    
\end{multline}
The Fourier transform of this can now be evaluated easily and gives us (\ref{eq:Fan_EoM_qubit_In_matrix_FT}). 

A similar set of steps allows us to also determine that
\begin{multline}
     \langle 0 |\hat{a}_{c}(t) \hat{a}_{\mathrm{in},p}(t) |u_1 u_2^{+} \rangle_{d_1,d_2} 
    \\ = \frac{1}{2\pi}  \big[S_{c,d_2}(u_2)  \delta_{p,d_1} + S_{c,d_1}(u_1) \delta_{p,d_2} \big]  e^{-i (u_1 + u_2) t}
\label{eq:Fan_EoM_c_portIn_matrix_time}    
\end{multline}
by recognizing from (\ref{eq:Fan_EoM_c_input_output_QIOT_matrix_Fourier_result}) that
\begin{align}
    \langle 0 |\hat{a}_{c}(t)|u^{+} \rangle_{p} = \frac{1}{\sqrt{2 \pi}} S_{c,p}(u) e^{-iut}.
\end{align}
The Fourier transform of (\ref{eq:Fan_EoM_c_portIn_matrix_time}) then gives us (\ref{eq:Fan_EoM_c_portIn_matrix_FT}). 
\section{Correlated Scattering Amplitude}
\label{sec:QIOT_TwoPhoton_connected_scattering}
Here, we derive the Fourier transform of 
\begin{align*}
    {}_{b}\langle v_1^{+} |\hat{\sigma}_{+}(t) |0\rangle \langle 0|\hat{\sigma}_{-}(t)\hat{a}_c(t)|u_1 u_2^{+} \rangle_{d_1,d_2}
\end{align*}
that arises in (\ref{eq:Fan_EoM_qubit_lowering_operator_matrix_TwoPhotonTransport_2}) for the two-photon transport analysis. This term is related to the correlated scattering effects in the cavity-qubit system, which explicitly accounts for how the presence of the qubit can cause significant deviations between the single- and two-photon transport properties. From basic Fourier theory, we recognize that to evaluate this Fourier transform will require us to compute the convolution of the Fourier transform of the two matrix elements involved. The Fourier transform of the second matrix element has already been solved for in (\ref{eq:Fan_EoM_qubitc_lowering_operator_matrix_TwoPhotonTransport_Result}); hence, we need to begin here by evaluating the Fourier transform of ${}_{b}\langle v_1^{+} |\hat{\sigma}_{+}(t) |0\rangle$.

Using the single-photon transport solution from Section \ref{sec:OnePhotonScattering} and following similar steps as in Appendix \ref{sec:TwoPhoton_input}, we find that 
\begin{align}
    {}_{b}\langle v_1^{+} |\hat{\sigma}_{+}(t) |0\rangle 
     = (\langle 0 |\hat{\sigma}_{-}(t) |v_1^{+} \rangle_{b})^{\dagger} = \frac{1}{\sqrt{2\pi}} S^{*}_{q,b}(v_1) e^{i v_1 t}.
    \label{eq:Fan_EoM_qubit_lowering_operator_matrix_assump_Fourier_result_time_conj}
\end{align}
Then, taking the Fourier transform, we obtain
\begin{align}
    {}_{b}\langle v_1^{+} |\hat{\sigma}_{+}(v_2) |0\rangle = S^{*}_{q,b}(v_1) \delta(v_1 + v_2).
    \label{eq:Fan_EoM_qubit_lowering_operator_matrix_assump_Fourier_result_time_conj_FT}
\end{align}
Next, the Fourier transform of the second matrix element is
\begin{multline}
     \langle 0|\hat{\sigma}_{-}(v_2) * \hat{a}_c(v_2)|u_1 u_2^{+} \rangle_{d_1,d_2}    = \Gamma \big({|u_1 u_2^{+} \rangle_{d_1,d_2}},v_2 \big) \times \\ \delta({ v_2 - (u_1 + u_2) }),
     \label{eq:Fan_EoM_qubitc_lowering_operator_matrix_TwoPhotonTransport_Result_SimpleNotation}
\end{multline}
where $\Gamma$ has been given in (\ref{eq:Fan_EoM_qubitc_lowering_operator_matrix_TwoPhotonTransport_Alt_part1_FT_SimpleNotation_coeff}). The convolution of (\ref{eq:Fan_EoM_qubit_lowering_operator_matrix_assump_Fourier_result_time_conj_FT}) and (\ref{eq:Fan_EoM_qubitc_lowering_operator_matrix_TwoPhotonTransport_Result_SimpleNotation}) can now be readily evaluated, yielding the result given in (\ref{eq:Fan_EoM_qubitc_lowering_operator_matrix_TwoPhotonTransport_Alt_part1_FT}).

\section{Two-Photon Transport Expressions}
\label{sec:Smat_general_case}
In this appendix, we provide more details on the derivation of (\ref{eq:ScatteringMatrix_TwoPhotons_explicit_main_ScatteringCase2}) and (\ref{eq:ScatteringMatrix_TwoPhotons_explicit_main_ScatteringCase4}) as representative cases to illustrate the general process. As noted in the main text, the compact nature of the final expressions here comes for the case where we assume the ports are symmetric such that $\kappa_1 = \kappa_2 = \kappa$.

For the case of (\ref{eq:ScatteringMatrix_TwoPhotons_explicit_main_ScatteringCase2}), we are considering when the photon of frequency $v_1$ exits Port $2$ and the photon of frequency $v_2$ exits Port $1$. We then have from (\ref{eq:ScatteringMatrix_TwoPhotons_ResIden_2}) that
\begin{multline}
{}_{2,1}\langle v_1 v_2^{-}| u_1 u_2^{+}\rangle_{1,1} 
    = t_{21}(v_1) \big[ 1
    -i \alpha  \kappa \big] \\ \times [\delta(u_1 - v_1) \delta(u_2 - v_2)  + \delta(u_2 - v_1)  \delta(u_1 - v_2) ]
    \\ + i\sqrt{\kappa} \, \big[t_{21}(v_1) \, \beta_{1} + r_2(v_1) \, \beta_{2} \big] \delta\big((v_1 + v_2) - (u_1 + u_2) \big),
    \label{eq:app_scattering_2photon_1}
\end{multline}
where
\begin{align}
\alpha = \frac{v_2 - \omega_q}{\big[ v_2 -\omega_c +i \kappa \big](v_2 - \omega_q)  -g^2},
\label{eq:Alpha_shofthand_notation_appendix}
\end{align}
\begin{multline}
\beta_{p}  = \frac{  1}{\big[ v_2 -\omega_c +i \kappa  \big](v_2 - \omega_q) -g^2} 
    \\ \times \frac{g^2}{\pi} S^{*}_{q,p}(v_1) \Gamma \big({|u_1 u_2^{+} \rangle_{1,1}},u_1 + u_2 \big).
\label{eq:Beta_shofthand_notation_appendix}
\end{multline}
Noting that $ 1 -i \alpha \kappa  = r_1(v_2)$, we get that
\begin{multline}
{}_{2,1}\langle v_1 v_2^{-}| u_1 u_2^{+}\rangle_{1,1}  = t_{21}(v_1) r_1(v_2) \\ \times [\delta(u_1 - v_1) \delta(u_2 - v_2)  + \delta(u_2 - v_1)  \delta(u_1 - v_2) ]
    \\ + \frac{i g }{\pi} S^{}_{q,2}(v_1) S^{}_{q,1}(v_2) \Gamma \big({|u_1 u_2^{+} \rangle_{1,1}},u_1+u_2 \big) \times \\ \delta\big((v_1 + v_2) - (u_1 + u_2) \big), 
\label{eq:ScatteringMatrix_TwoPhotons_explicit_main_ScatteringCase2_appendix}
\end{multline}
where we have also recombined terms in the last line of (\ref{eq:app_scattering_2photon_1}) into a consolidated form.

For the case of (\ref{eq:ScatteringMatrix_TwoPhotons_explicit_main_ScatteringCase4}), we now have the photon of frequency $v_1$ exits Port $1$ and the photon of frequency $v_2$ exits Port $2$. Here, (\ref{eq:ScatteringMatrix_TwoPhotons_ResIden_2}) becomes
\begin{multline}
{}_{1,2}\langle v_1 v_2^{-}| u_1 u_2^{+}\rangle_{1,1} 
    = r_1(v_1) ( -i\alpha  \kappa  )
    \\ \times [\delta(u_1 - v_1) \delta(u_2 - v_2)  + \delta(u_2 - v_1)  \delta(u_1 - v_2) ]
    \\ + i\sqrt{\kappa} \, \big[r_1(v_1) \, \beta_{1} + t_{12}(v_1) \, \beta_{2} \big] \,\, \delta\big((v_1 + v_2) - (u_1 + u_2) \big).    
    \label{eq:app_scattering_2photon_2}
\end{multline}
In this case, we can find that $-i\alpha\kappa = t_{21}(v_2)$, which when combined with the same kind of consolidation that led to (\ref{eq:ScatteringMatrix_TwoPhotons_explicit_main_ScatteringCase2_appendix}) for the last line of (\ref{eq:app_scattering_2photon_2}) gives us (\ref{eq:ScatteringMatrix_TwoPhotons_explicit_main_ScatteringCase4}). Similar kinds of manipulations also allow one to derive (\ref{eq:ScatteringMatrix_TwoPhotons_explicit_main_ScatteringCase1}) and (\ref{eq:ScatteringMatrix_TwoPhotons_explicit_main_ScatteringCase3}).

\section{Correlated Scattering Contour Integral}
\label{sec:Contour_Integral}
Here, we show how to evaluate the correlated scattering term in (\ref{eq:general_output_state_FT_rr2}) given by
\begin{multline}
H(x_1,x_2) = \frac{1}{2} \iint_{-\infty}^{\infty} \mathrm{d} v_1 \mathrm{d} v_2 P(v_1,v_2;x_1,x_2) \\ \times \frac{i g }{\pi} S^{}_{q,b_1}(v_1) S^{}_{q,b_2}(v_2) \Gamma \big({|u_1 u_2^{+} \rangle_{1,1}},u_1+u_2 \big) \\ \times  \delta\big((v_1 + v_2) - (u_1 + u_2) \big). 
\label{eq:Contour_Integral1}
\end{multline}
As mentioned in the main text, this is most easily done through a convenient change of variables. In particular, we can define $E_i = u_1 + u_2$, $E_o = v_1 + v_2$, $\Delta_i = (u_1 - u_2)/2$, $\Delta_o = (v_1 - v_2)/2$, $x_m = x_1 - x_2$, and $x_c = (x_1 + x_2)/2$. In terms of these new variables, we have that
\begin{align}
P(v_1,v_2;x_1,x_2) =  \frac{1}{\pi\sqrt{2}} e^{i E_o x_c /v_g } \cos(\Delta_o x_m/v_g),
\label{eq:Contour_Integral2}
\end{align}
\begin{multline}
S^{}_{q,b_1}(v_1) S^{}_{q,b_2}(v_2) =  \\
 \frac{\sqrt{\kappa_{}} g}{(\Delta_o + \tfrac{E_o}{2} - \lambda_{1,-}) (\Delta_o + \tfrac{E_o}{2} - \lambda_{1,+}) }    \\ \times \frac{\sqrt{\kappa_{}} g}{(\Delta_o - \tfrac{E_o}{2} + \lambda_{1,-}) (\Delta_o - \tfrac{E_o}{2} +  \lambda_{1,+}) } .
\label{eq:Contour_Integral3}
\end{multline}
Adjusting the integration variables accordingly, we then have that (\ref{eq:Contour_Integral1}) becomes
\begin{multline}
H(x_1,x_2) = \frac{1}{\sqrt{2}2\pi} \Gamma \big({|u_1 u_2^{+} \rangle_{1,1}},E_i \big)  \int_{-\infty}^{\infty} \mathrm{d} E_o \int_{-\infty}^\infty \mathrm{d} \Delta_o \\ \times e^{i E_o x_c /v_g } \cos(\Delta_o x_m/v_g)  \frac{i g }{\pi} S^{}_{q,b_1}(v_1) S^{}_{q,b_2}(v_2)    \delta\big(E_o - E_i \big). 
\label{eq:Contour_Integral_variable_change}
\end{multline}
After evaluating the trivial $E_o$ integral, the $\Delta_o$ integral can be further rewritten to give us 
\begin{multline}
H(x_1,x_2) = \frac{i g }{\pi} \Gamma \big({|u_1 u_2^{+} \rangle_{1,1}},E_i \big)  \times   \\  \frac{\kappa g^2}{\sqrt{2} 4\pi}  e^{i E_i x_c/v_g}   \int_{-\infty}^{\infty}  \mathrm{d} \Delta_o  \\  \times \frac{ e^{i \Delta_o x_m/v_g} +e^{-i \Delta_o x_m/v_g} }{(\Delta_o + \tfrac{E_i}{2} - \lambda_{1,-}) (\Delta_o + \tfrac{E_i}{2} - \lambda_{1,+})  } \\ \times \frac{1}{(\Delta_o - \tfrac{E_i}{2} + \lambda_{1,-}) (\Delta_o - \tfrac{E_i}{2} +  \lambda_{1,+}) }. 
\label{eq:Contour_Integral5}
\end{multline}

This final integral over $\Delta_o$ can be evaluated using standard contour integral techniques; e.g., as detailed in \cite{stone2009mathematics}. The sign of $x_m$ will dictate how the contour needs to be closed (i.e., in the upper or lower half-plane) for the different exponential terms; however, due to how the poles are located and the form of the integral the results of the pole residues are the same regardless. Correspondingly, we can find that the result of the integral in (\ref{eq:Contour_Integral5}) can be compactly written as 
\begin{multline}
\frac{4\pi i}{( \lambda_{1,+}  - \lambda_{1,-}) ({E_i} - \lambda_{1,+} - \lambda_{1,-})} \times \\ \Bigg[ \frac{ e^{i ({E_i} - 2\lambda_{1,-}) \tfrac{|x_m|}{2}} }{({E_i} - 2\lambda_{1,-})   }  
 - \frac{ e^{i ({E_i} - 2\lambda_{1,+}) \tfrac{|x_m|}{2} }}{({E_i} - 2\lambda_{1,+})  }  \Bigg],
\label{eq:Contour_Integral8}    
\end{multline}
where the $4\pi i$ comes from the fact that there are the two seperate complex exponentials in (\ref{eq:Contour_Integral5}) that are being integrated. Substituting the above into (\ref{eq:Contour_Integral5}), we obtain the final expression as shown in (\ref{eq:H_general_final_result}) of the main text.

\acknowledgements{This research was sponsored by the National Science Foundation under Grant No. 2202389.}

\bibliographystyle{emsreport}
\bibliography{main}

\iftrue 

\fi 

\end{document}